\documentclass{lmcs}
\pdfoutput=1

\usepackage{lastpage}
\lmcsdoi{22}{4}{25}
\lmcsheading{}{\pageref{LastPage}}{}{}%
{Aug.~29,~2025}{Dec.~01,~2026}{}


\usepackage{mathtools}
\usepackage{amsthm}
\usepackage{amssymb}
\usepackage{tikz}
\usepackage{graphicx}
\usepackage{booktabs}
\usepackage{xcolor}
\usepackage{algorithm}
\usepackage{microtype}
\usepackage[noend]{algpseudocode}
\usepackage{xspace}
\usepackage{url}
\usepackage{pgfplots} 
\usepackage{pgfplotstable} 
\pgfplotsset{compat=1.18}

\usepackage{iftex}

\ifpdf
  \usepackage{underscore}         
  \usepackage[english]{babel}
  \usepackage[utf8]{inputenc}     
  \usepackage[T1]{fontenc}        
\else
  \usepackage{breakurl}           
\fi
\usetikzlibrary{automata,positioning}

\newcommand{\Nat}{\mathbb{N}}
\newcommand{\Lang}{\mathcal{L}}

\newcommand{\mfalse}{\mathtt{false}}
\newcommand{\mtrue}{\mathtt{true}}
\newcommand{\Fib}{\mathtt{Fib}} 
\newcommand{\memory}{\mathtt{memory}} 
\newcommand{\C}{\mathtt{C}} 
\newcommand{\B}{\mathtt{B}}
\newcommand{\fib}{\mathit{fib}}

\def\CC{{C\nolinebreak[4]\hspace{-.05em}\raisebox{.4ex}{\tiny ++}}\xspace}

\algrenewcommand\algorithmicrequire{\textbf{Input:}}
\algrenewcommand\algorithmicensure{\textbf{Output:}}

\algblockdefx[PARNAME]{ParDo}{EndParDo}%
[1]{\textbf{do in parallel for } #1}%
{}
\makeatletter
\ifthenelse{\equal{\ALG@noend}{t}}%
  {\algtext*{EndParDo}}
  {}%
\makeatother

\newcommand{\GPUexplore}{\textsc{GPUexplore}\xspace}
\newcommand{\SLCO}{\textsc{Slco}\xspace}

\usepackage{listings}

\definecolor{frenchplum}{RGB}{129,20,83}
\definecolor{deepfuchsia}{rgb}{0.76, 0.33, 0.76}
\definecolor{ferrarired}{rgb}{1.0, 0.11, 0.0}
\definecolor{darkcandyapplered}{rgb}{0.64, 0.0, 0.0}

\lstdefinelanguage{slco}{
	columns=fullflexible,
	morekeywords={Byte,Integer,Boolean,false,true},
    	keywordstyle={\color{blue}},
	classoffset=1,
	morekeywords = {model, classes, variables, state, machines, objects, initial, forbidden, states, transitions, channels, ports, actions},
	keywordstyle=\color{frenchplum},
	classoffset=2,
        morekeywords={sync,between,and,or,not},
    	keywordstyle=\color{ferrarired},
        classoffset=3,
        morekeywords={send,receive,to,from,after,ms,do,comm},
        keywordstyle=\color{deepfuchsia},
        classoffset=4
}

\def\slcoinline{\lstinline[language=slco, columns=fixed]}

\begin{document}

\title[Parallel DFA Minimisation, Equivalence Checking and Inclusion Checking]{Evaluating Massively Parallel Algorithms for DFA Minimisation, Equivalence Checking and Inclusion Checking}

\specialissue{Selected Papers of the 15th International Symposium on Games, Automata, Logics, and Formal Verification (GandALF 2024)}

\author[J. Heemstra]{Jan Heemstra\lmcsorcid{0009-0007-2025-5085}}[a]
\author[J. Martens]{Jan Martens\lmcsorcid{0000-0003-4797-7735}}[b]
\author[A. Wijs]{Anton Wijs\lmcsorcid{0000-0002-2071-9624}}[a]

\address{Eindhoven University of Technology, The Netherlands} 
\email{j.h.heemstra@tue.nl, a.j.wijs@tue.nl}
\thanks{This work is supported by NWO grant OCENW.M.21.061 for the GAP project.}

\address{Leiden University, The Netherlands}
\email{j.j.m.martens@liacs.leidenuniv.nl}

\begin{abstract}
We study parallel algorithms for the minimisation and equivalence checking of Deterministic Finite
Automata~(DFAs). Regarding DFA minimisation, we implement four different massively parallel
algorithms on Graphics Processing Units~(GPUs). Our results confirm the
expectations that the algorithm with the theoretically best time complexity is
not practically suitable to run on GPUs due to the large amount of resources
needed. We empirically verify that parallel partition refinement algorithms from
the literature perform better in practice, even though their time complexity is
worse. Furthermore, we introduce a novel algorithm based on partition refinement with
an extra parallel partial transitive closure step and show that on specific
benchmarks it has better run-time complexity and performs better in practice.    

In addition, we address checking the language equivalence and inclusion of two DFAs.
We consider the Hopcroft-Karp 
algorithm, and explain how a variant of it can be parallelised for GPUs. We note that these problems can be 
encoded for the GPU-accelerated model checker \GPUexplore, allowing to use its lockless hash table and fine-grained parallel work distribution mechanism.
\end{abstract}

\maketitle

\section{Introduction}
In contrast to sequential chips, the processing power of parallel devices keeps
increasing. Graphics Processing Units, or GPUs, are examples of such devices.
Originating from the need to do simple computations for many (independent)
pixels to generate graphics, GPUs have also shown useful as computational
powerhouses, and led to general-purpose computing on GPUs~(GPGPU). Most
convincingly, GPUs have become indispensable in training models for artificial
intelligence. Because of the enormous potential of GPUs, it is important to
investigate how computational problem solving can be accelerated with them.

Deterministic Finite Automata~(DFAs) are one of the simplest computational
formalisms. The natural problem of computing a minimal machine that is
equivalent to a given machine w.r.t.\ the input is omnipresent in the field of theoretical computer
science. In the case of DFAs the problem has a rich history. The first method
that computes a minimal DFA dates back to Moore's
framework~\cite{moore1956gedanken}, and is a \emph{partition refinement}
algorithm. Later, this algorithm was adapted by
Hopcroft~\cite{hopcroft1971DFAmin} to a quasi-linear time algorithm.

The complexity class known as Nick's Class~($NC$) consists of the problems that
can be solved in polylogarithmic time with a parallel
machine using a polynomial number of parallel processors. It is an open question
whether \smash{$\textit{NC} \stackrel{?}{=} \textit{P}$}, but it is widely believed that this is not the case.
Similar to the assumption that
 decision problems not in $\textit{P}$
  are inherently
difficult~(known as Cobham's thesis), we can think of $P$-complete problems as
being inherently sequential. 

The problem of minimising DFAs is known to be in
\textit{NC}~\cite{cho1992parallelcoarsestset}, which intuitively means it can be
efficiently computed in parallel. In contrast, the problem of computing
bisimilarity on non-deterministic structures is known to be
$P$-complete~\cite{balcazar1992}. Interestingly, the most efficient sequential
algorithms for these two problems, i.e., Hopcroft's
algorithm~\cite{hopcroft1971DFAmin} and an algorithm based on
Paige-Tarjan~\cite{paige1987three}, respectively, which formed the basis for other
algorithms computing weak variants of bisimilarity~\cite{GroWij16,GJKW17,JGKW20}, are very similar. In
particular, these algorithms are both \emph{partition refinement} algorithms.  

The parallel algorithms studied for computing 
bisimilarity on non-deterministic structures
 and DFA minimisation are also partition refinement
algorithms~\cite{pram-bisim-facs,martens2022linear,ravikumar1996paralleldfa,
tewari2002paralleldfa,wijs2015}. Since DFA minimisation is in $\textit{NC}$,
there is a parallel sublinear time algorithm. However, none of these partition
refinement algorithms studied have a sublinear run-time. A linear lower bound
for the parallel run-time was proven in~\cite{groote2023lowerbounds} for any
parallel partition refinement algorithm deciding bisimilarity, and this result
also directly applies to deterministic structures such as DFA minimisation. This
means that no partition refinement algorithm can achieve the theoretically
optimal run-time on parallel machines.  
It is therefore interesting to investigate whether there is an algorithm that is not a
partition refinement algorithm that performs better in parallel than partition refinement algorithms.  
  
The algorithm introduced in~\cite{cho1992parallelcoarsestset} runs in
logarithmic time. However, the work is mainly theoretical and the large number
of parallel processors and the large amount of memory required makes it unlikely to scale well in
practice. The main constraint here is the need to compute the transitive closure for
the underlying graph of the DFA. It seems hard to find a significant improvement
in the number of parallel processors needed. 

In this article, we compare implementations of different parallel algorithms
for DFA minimisation on GPUs, using the various parallel algorithms proposed in the literature
as a basis. This comparison can be found in Section~\ref{sec:min}.
We establish that the logarithmic runtime
complexity with the construction from~\cite{cho1992parallelcoarsestset} is not
feasible due to the large number of processors needed. Additionally, we find
that on our benchmarks, described in Section~\ref{sec:exps},
 the partition refinement algorithm that uses sorting in
each iteration performs better than the naive splitting strategy on the more
diverse benchmarks from the VLTS benchmark set,\footnote{https://cadp.inria.fr/resources/vlts (visited on: 07-07-2026).} but worse for benchmarks that are
known to be hard for partition refinement algorithms. Finally, we show a method
of adding a partial transitive closure as a preprocessing step that can
significantly increase the performance on benchmarks with a very specific shape.

This article is based on a previously published paper~\cite{gandalf24}. While the work
on DFA minimisation stems from that earlier publication, the current article extends the scope
by also considering equivalence checking and inclusion checking of DFAs,
two other fundamental operations
that have their practical use in applications such as model checking and compiler construction.
A well-known algorithm for DFA equivalence checking is the algorithm by Hopcroft and
Karp~\cite{hopcroft-karp,AhHoUl74}. Its almost linear complexity is due to the fact that the
algorithm constructs the transitive closure of a bisimulation relation using a union-find
data structure. Although this is hard to achieve in a massively parallel way, we show that a
naive version of the algorithm, in which the bisimulation relation is constructed using a hash
table, but its transitive closure is not explicitly maintained, works well in practice.
To achieve this, in Section~\ref{sec:dfa-eq}, we use the GPU-accelerated explicit-state model checker 
\GPUexplore~\cite{gpuexplore3,gpuexplore3-journal},
which contains all the ingredients to encode the problem of DFA equivalence checking, and
perform the computations on the GPU. To demonstrate the effectiveness of this approach,
we define some additional benchmarks, and report on the results obtained with these benchmarks,
 in Section~\ref{sec:exps}. The naive version of the algorithm by Hopcroft and Karp can also be
 used for inclusion checking, i.e., checking that the language of one DFA is included in the 
 language of another DFA. Also for this operation, we explain how to encode this for \GPUexplore, perform experiments and discuss the performance of our approach.
 
 Finally, in Section~\ref{sec:conc}, we draw conclusions and
 provide pointers for future work.

\section{Preliminaries}
We write $\mathbb{B} = \{\mtrue, \mfalse\}$ for the set of booleans, $\Nat$ for the set
of natural numbers, and for numbers $i,j\in \Nat$ we define $[i,j] = \{ c \in
\Nat \mid i \leq c \leq j\} \subseteq \Nat$, the closed interval from $i$ to
$j$. Given an alphabet $\Sigma$, a sequence $a_1 a_2 \ldots a_n$ of symbols from
$\Sigma$ is called a word. We write $\Sigma^*$ for the set containing all finite
sequences of letters in $\Sigma$. The empty sequence consisting of no symbols is
written as $\varepsilon$. 

\begin{defi}{(Deterministic Finite Automaton)}
  A deterministic finite automaton~(DFA) $A = (Q, \Sigma, \delta, F, q_0)$ is a five-tuple consisting of:
  \begin{itemize}\setlength\itemsep{0em}
    \item a finite set of states $Q$,
    \item a finite alphabet $\Sigma$, 
    \item a transition function $\delta: Q\times \Sigma \to Q$, 
    \item a set of accepting states $F\subseteq Q$, and
    \item an initial state $q_0 \in Q$. 
  \end{itemize}
\end{defi}

\noindent
We sometimes write $q \xrightarrow{a} q'$ if $\delta(q, a) = q'$. The function
$\delta^*: Q \times \Sigma^* \to Q$ extends the transition function to words and
is defined inductively for all words in $\Sigma^*$ as follows:
\begin{align*}
  \delta^*(q,\varepsilon) &= q \\
  \delta^*(q,aw) &= \delta^*(\delta(q,a), w)
\end{align*}

\noindent
Given a DFA $A=(Q,\Sigma,\delta, F,q_0)$, a word $w \in \Sigma^*$ is
 \emph{accepted} iff $\delta^*(q_0, w) \in F$. The language of a DFA $A$,
 notation $\Lang(A)$, is the set of all words $w \in \Sigma^*$ that are accepted
 by $A$. 

We consider the problem of computing the minimal DFA, i.e., given a DFA $A$,
identifying the DFA $A'$ with the smallest number of states such that $\Lang(A')
= \Lang(A)$. 

Minimising DFAs consists of combining indistinguishable states and deleting
unreachable states. The main part of the problem consists of combining states, and
removing unreachable states can be seen as a simple pre-processing step. For
the remainder of the article, we assume that all the states in a DFA are reachable
from its initial state. The
algorithms can be seen as computing bisimilarity, or the coarsest set partition
problem, without the preprocessing step that removes unreachable states.

\paragraph*{Representation.} For an input
automaton $A= (Q, \Sigma, \delta, F, q_0)$, we assume that the states in $Q$ and letters in
$\Sigma$ are represented by unique indices, i.e., $Q = \{0, \dots, \lvert Q
\rvert-1\}$ and $\Sigma = \{0, \dots \lvert \Sigma \rvert-1\}$. The transition
function $\delta$ is represented by $\lvert \Sigma \rvert$ arrays of length
$\lvert Q \rvert$, such that for state $q\in Q$ and letter $a\in \Sigma$, $\delta[a][q] = \delta(q,a)$.  

\paragraph*{The PRAM Model.} The complexities we mention assume the model of
the \emph{Parallel Random Access Machine~(PRAM)}. The PRAM is a natural extension
of the RAM model, where parallel processors have access to a shared memory. A PRAM consists of a sequence of processors 
$P_0, P_1, \dots$ and a function $\mathcal{P}$ that, given the size of the input, defines a bound on the number of processors used. 

Each processor $P_i$ has the natural instructions of a normal RAM and, in addition, has an instruction to
retrieve its unique index $i$. All processors run the same program in lock-step, using their index to identify the data they need to access.
This parallel processing is called single-instruction multiple data~(SIMD).  

There are many different ways to handle data-races. 
We assume the concurrent read, concurrent write~(CRCW) model following~\cite{stockmeyer1984simulation}, 
where processors are allowed to read from and write to the same memory location concurrently.
After multiple concurrent writes to the same memory location, that location contains the result
of one of those writes.

\paragraph*{GPUs.}
While in reality no device completely adheres to the PRAM model, recent hardware advancements have led
to devices that are getting better and better at approximating this model. The GPU, in particular, is a very suitable
target platform for PRAM algorithms, as it has been specifically designed for SIMD processing.
The performance of GPU programs typically relies on launching tens to hundreds
of thousands of threads, as the performance of these programs is often memory-bound: accessing the input data
in the GPU's \emph{global} memory, in NVIDIA CUDA terminology, is relatively slow. This latency can be hidden
by a GPU via fast context switching between threads. As one thread is waiting for data to be retrieved, another thread
is executed in the meantime on the same processor. It is this fast context switching between threads that allows GPUs,
typically equipped with several thousands of cores, to virtually execute hundreds of thousands of threads concurrently.
In the current work, we employ NVIDIA GPUs. Programs for these devices can be written in CUDA \CC.

\section{Algorithms for DFA minimisation}
\label{sec:min}
\subsection{Transitive closure} 
The first algorithm we discuss for DFA minimisation has theoretical polylogarithmic
runtime~\cite{cho1992parallelcoarsestset}. However, the large amount of memory
and parallel processors it uses makes it unlikely to work in practice. Here we
confirm this fact.  

The idea is rather simple; build a graph with nodes $V = Q\times Q$ and edges
$E$ containing $(q, q') \rightarrow (p, p')$ iff there is a letter $a\in \Sigma$
such that $\delta(q,a) = p$ and $\delta(q', a) = p'$. Initially, in the array
$\mathit{Apart}$ we label the nodes $(q,q') \in V$ to be inequivalent if $q
\in F \iff q' \not \in F$. Any two states $q, q' \in Q$ are not equivalent
iff they were initially labelled in $\mathit{Apart}$ or there is a path to a
labelled node. Computing this reachability of false nodes can be seen as
computing the transitive closure in the directed graph $(V,E$) containing $n^2$
nodes. In parallel this computation can be done in polylogarithmic running time
using $O(\lvert V\rvert^3)$ parallel processors~\cite[Chapter 5.5.]{jaja1992introduction}. 

Algorithm~\ref{alg:fulltrans}, which we refer to as \texttt{trans},
 implements this idea. First, at lines~4--7 (l.4--7), the
graph is constructed, with inequivalent nodes labelled in the $\mathit{Apart}$ data
structure and the edges stored in the adjacency matrix $\mathit{Reach}$. Next,
the parallel transitive closure of $\mathit{Reach}$ is computed and
$\mathit{Apart}$ updated accordingly. If in an iteration there is no new pair of
states labelled $\mathit{Apart}$ the algorithm is finished. The minimal automaton
is represented in the graph where states $q, q' \in Q$ can be combined if $\neg
\mathit{Apart}(q,q')$.

Computing the transitive closure for a directed graph in logarithmic time
requires many processors. Our naive implementation requires $\lvert V\rvert^3$
parallel processors. Given a DFA with $n$ states, this means that since
$\lvert V \rvert = n^2$, we require $n^6$ processors. Theoretically, more
efficient methods are known for computing the transitive closure that use
matrix multiplication. Matrix multiplication can be computed with
$O(n^\omega)$ operations, where currently $\omega \leq 2.372\dots$, this
means we can compute our transitive closure with $O(\lvert V \rvert^\omega)$
processors. Since these algorithms are non-trivial and already $\lvert V \rvert
= n^2$, we believe these improvements would not significantly change the results
mentioned here. Relative to the other algorithms discussed in this article, the
required number of processors would still be very high.

\begin{algorithm}[tb]
  \begin{algorithmic}[1]
    \Require{A DFA $A = (Q, \Sigma, \delta, F, q_0)$ where $\lvert Q \rvert = n$}
    \Ensure{The minimal quotient automaton represented in the matrix $\mathit{Apart}$}
    \State $V :: Q\times Q$ \Comment{Nodes of graph consisting of pair of states}
    \State $\mathit{Apart} ::$ Array$[n^2]$ of type $\mathbb{B}$ 
    \State $\mathit{Reach} ::$ Array$[n^2][n^2]$ of type $\mathbb{B}$ \Comment{Represents reachability in $V$, initially $\mfalse$}
    \ParDo{$(q, q')\in V$ }
      \Comment{Initialises data structures in parallel.}
      \State $\mathit{Apart}[(q,q')] := (q \in F \iff q'\not\in F)$ \Comment{State initially unequal}
      \ForAll{$a\in \Sigma$}
        \State $\mathit{Reach}[(q, q')][(\delta(q,a), \delta(q',a))] := \mtrue$
      \EndFor
    \EndParDo
    \State $\mathit{stable} := \mfalse$
    \While{$\neg \mathit{stable}$} 
      \State $\mathit{stable} := \mtrue$
       \ParDo {$s,t,u \in V$ }
        \If{$\mathit{Reach}[s][t] \text{ and } \mathit{Reach}[t][u] \text{ and } \neg \mathit{Reach}[s][u]$}
          \State $\mathit{Reach}[s][u] := \mtrue$
        \EndIf
        \EndParDo
      \ParDo {$s,t \in V$}
        \If{$\mathit{Reach}[s][t] \text{ and } \mathit{Apart}[t] \text{ and } \neg \mathit{Apart}[s]$}
          \State $\mathit{Apart}[s] := \mtrue$
          \State $\mathit{stable} := \mfalse$
        \EndIf
      \EndParDo
    \EndWhile 
\end{algorithmic}
\caption{\label{alg:fulltrans} Transitive DFA minimisation~$\mathtt{trans}$.}
\end{algorithm}

\subsection{Naive partition refinement}
The next algorithm for DFA minimisation, \texttt{naivePR},
 is an adaptation of the parallel algorithm for bisimilarity
checking of Labelled Transition Systems (LTSs) from~\cite{martens2022linear}.
The program runs on a PRAM with $m$ processes, where $m = \lvert \Sigma \rvert * n$ is the number of transitions in the
input DFA, and $n$ is the number of states.
 
The algorithm applies \emph{partition refinement}: states are initially partitioned into \emph{blocks},
and the algorithm repeatedly splits blocks into smaller blocks until a fix-point is reached.
Once this has happened, each block represents one state of the minimised DFA.

When splitting blocks in parallel, one particular challenge is how to identify
newly created blocks, as each new block requires a unique identifier.
Algorithm~\ref{alg:partref} does this by means of a leader election procedure:
for each block, one of its states is elected leader, meaning that it is used as
an identifier to refer to the block. In this way each iteration of the algorithm
takes constant time if performed on a parallel machine that has concurrent
writes.

In Algorithm~\ref{alg:partref}, at l.\ref{l:partref-block}, an array
\textit{block} is initialised in such a way that it defines for every state in $Q$ its current
block (as represented by a leader in $Q$). An array \textit{new\_leader} is
defined at l.\ref{l:partref-newleader} that is used to elect new leaders. At
l.\ref{l:partref-initleaders}, the initial leaders are selected: one state $q_f
\in F$ for the block consisting of all the accepting states $q \in F$, and one
state $q_n \in Q \setminus F$ for all the non-accepting states $q' \in Q
\setminus F$. The array \textit{block} is subsequently initialised using these
leaders (l.\ref{l:partref-for-initblock}--\ref{l:partref-initblock}).

Next, partition refinement is applied inside the \textbf{while}-loop at l.\ref{l:partref-whileneqstable}. The variable
\textit{stable} is used to monitor whether a fix-point has been reached, which happens as soon as no blocks
can be split any further. At the start of each iteration through the \textbf{while}-loop, \textit{stable} is set to
$\mtrue$ (l.\ref{l:partref-setstable}). Next, all transitions of the DFA are processed in parallel (l.\ref{l:partref-leaderfor}),
and for each transition $q \xrightarrow{a} q'$, it is checked whether $\mathit{block}[q']$ differs from the block that
the leader $\mathit{block}[q]$ can reach via an $a$-transition. If it does, then $q$ should be separated from its
leader. At l.\ref{l:partref-leaderelect}, $q$ is assigned to $\mathit{new\_leader}[\mathit{block}[q]]$, the latter being the
position where the leader for the new block will be elected. Here, the result of concurrent writes, as allowed by the
PRAM CRCW model, is used for leader election.

Subsequently, when l.\ref{l:partref-getleaderfor} is reached, \textit{new\_leader} contains the newly elected leaders:
specifically, at $\mathit{new\_leader}[\mathit{block}[q]]$, the leader for the new block created by splitting off states from
$\mathit{block}[q]$ is stored. In the parallel loop of l.\ref{l:partref-getleaderfor}, the transitions are once more processed
in parallel, and whenever a state turns out to differ from its leader regarding block reachability (l.\ref{l:partref-get-leaderdiff}),
the leader of that state is updated (l.\ref{l:partref-getnewleader}). Finally, since a block has been split, \textit{stable}
is set to $\mfalse$ at l.\ref{l:partref-notstable}.
 
The largest difference between Algorithm~\ref{alg:partref} and the original
algorithm~\cite{martens2022linear} is that Algorithm~\ref{alg:partref} splits
blocks directly w.r.t.\ the leader, as opposed to first selecting one particular
block as \emph{splitter}, and splitting those blocks in which some states differ
w.r.t.\ their leader concerning the ability to reach the splitter. The reason
for this difference is that for DFAs, comparing the outgoing transitions of two
states is much more straightforward, as each state has exactly one outgoing
transition for every $a \in \Sigma$. In the setting of LTSs, due to
non-determinism it is not straightforward to directly compare the behaviour of a state
with the leader state, and hence a splitter is chosen to simplify this comparison. 

\begin{algorithm}
  \begin{algorithmic}[1]
    \Require{A DFA $A = (Q, \Sigma, \delta, F, q_0)$ where $\lvert Q \rvert = n$}
    \Ensure{The minimal quotient automaton represented in the array $block$}
    \State $\mathit{block} ::$ Array$[n]$ of type $Q$ \label{l:partref-block}
    \State $\mathit{new\_leader} ::$ Array$[n]$ of type $Q$ \label{l:partref-newleader}
    \State Select initial leader states~$q_f \in F$ and $q_{n}\in Q\setminus F$ \label{l:partref-initleaders}
    \ParDo{$q\in Q$} \label{l:partref-for-initblock}
      \State $\mathit{block}[q] := (q \in F\ ?\ q_f : q_n)$ \Comment Initialise \label{l:partref-initblock}
    \EndParDo
    \State $\mathit{stable} := \mfalse$
    \While{$\neg \mathit{stable}$} \label{l:partref-whileneqstable}
      \State $\mathit{stable} := \mtrue$ \label{l:partref-setstable}
      \ParDo{$(q,a)\in Q \times \Sigma$} \label{l:partref-leaderfor}
        \If{$\mathit{block}[\delta(q,a)] \neq \mathit{block}[\delta(\mathit{block}[q],a)]$} \label{l:partref-leaderdiff}
          \State $\mathit{new\_leader[block[q]]} := q$ \Comment Leader election \label{l:partref-leaderelect}
        \EndIf
      \EndParDo
      \ParDo{$(q, a) \in Q \times \Sigma$} \label{l:partref-getleaderfor}
        \If{$\mathit{block}[\delta(q,a)] \neq \mathit{block}[\delta(\mathit{block}[q],a)]$} \label{l:partref-get-leaderdiff}
          \State $\mathit{block[q]} := \mathit{new\_leader}[{\mathit{block}[q]}]$ \Comment Split from leader \label{l:partref-getnewleader}
          \State $\mathit{stable} := \mfalse$ \label{l:partref-notstable}
        \EndIf
      \EndParDo
    \EndWhile
\end{algorithmic}
\caption{\label{alg:partref} Parallel leader-election-based algorithm~$\mathtt{naivePR}$.}
\end{algorithm}

\begin{algorithm}
  \begin{algorithmic}[1]
    \Require{A DFA $A = (Q, \Sigma, \delta, F, q_0)$, where $\lvert Q \rvert = n$}
    \Ensure{The minimal quotient automaton represented in the array $block$}
    \State $\mathit{block} ::$ Array$[n]$ of type $Q$ 
    \State $\mathit{new\_block} :: Q$
    \State $\mathit{leader} :: Q$
    \State $\mathit{new\_leader} ::$ Array$[n]$ of type $Q$
    \State Select initial leader states~$q_f \in F$ and $q_{n}\in Q\setminus F$
    \ParDo{$q\in Q$}
      \State $\mathit{new\_leader}[q] := \bot$ \Comment Initialise
      \State $\mathit{block}[q] := (q \in F\ ?\ q_f : q_n)$
    \EndParDo
    \While{$\neg \mathit{stable}$}
      \State $\mathit{stable} := \mtrue$
      \ParDo{$(q,a)\in Q \times \Sigma$} \label{l:partref-cas-leaderfor}
        \State $\mathit{leader} := \mathit{block}[q]$
        \If{$\mathit{block}[\delta(q,a)] \neq \mathit{block}[\delta(\mathit{leader},a)]$}
          \State $\{\mathit{new\_block} := \mathit{new\_leader[leader]};$ \Comment Leader election with CAS \label{l:partref-cas-casbegin}
          \State $\mathit{new\_leader[leader]} = \bot\ ?\ new\_leader[leader] := q\}$ \label{l:partref-cas-casend}
          \State $\mathit{block[q]} := \mathit{new\_block} = \bot\ ?\ q: \mathit{new\_block}$ \Comment Split from leader \label{l:partref-cas-split}
          \State $\mathit{stable} := \mfalse$ \label{l:partref-cas-notstable}
        \EndIf
      \EndParDo
      \ParDo{$q\in Q$} \label{l:partref-cas-resetbegin}
        \State $\mathit{new\_leader[q]} := \bot$ \label{l:partref-cas-resetend}
      \EndParDo
    \EndWhile
\end{algorithmic}
\caption{\label{alg:partref-cas} Parallel leader-election based $\mathtt{naivePR}$ with atomics.}
\end{algorithm}

\noindent
In Algorithm~\ref{alg:partref}, leader election is performed in two phases: in the first phase (l.\ref{l:partref-leaderfor}--\ref{l:partref-leaderelect}),
states are written to $\mathit{new\_leader}$ to elect new leaders, and the results are subsequently read at l.\ref{l:partref-getleaderfor}--\ref{l:partref-notstable}.
One could argue that this is inefficient, and that it would perhaps be better to combine these two phases. This is possible, but it requires the use of
atomic \textit{compare-and-swap}~(CAS) operations. This is illustrated in Algorithm~\ref{alg:partref-cas}.
In the single loop at l.\ref{l:partref-cas-leaderfor}--\ref{l:partref-cas-notstable}, new leaders are written to \emph{and} read from
$\mathit{new\_leader}$. At l.\ref{l:partref-cas-casbegin}--\ref{l:partref-cas-casend}, the use of a compare-and-swap operation
is described: in one atomic operation, the current value stored at $\mathit{new\_leader[leader]}$ is stored in $\textit{new\_block}$, and
it is checked whether $\mathit{new\_leader[leader]}$ is equal to the initial value $\bot$, and if it is, it is set to $q$.
Next, at l.\ref{l:partref-cas-split}, if \textit{new\_block} is equal to $\bot$, it means $q$ has been elected as leader. Otherwise,
\textit{new\_block} indicates which state is the new leader. Note that for this to work, after execution of the loop at l.\ref{l:partref-cas-leaderfor},
the values of \textit{new\_leader} have to be reset to $\bot$.

In practice, we experienced that a GPU implementation (in CUDA 12.2) of Algorithm~\ref{alg:partref-cas} exhibits similar runtimes compared to a GPU implementation
of Algorithm~\ref{alg:partref}. The benefit of merging the loops seems to be negated by the use of atomic operations. For this reason, when discussing
the experiments in Section~\ref{sec:results}, we do not involve Algorithm~\ref{alg:partref-cas}.

\subsection{Sorting arrays}
The next algorithm for DFA minimisation is an algorithm inspired by~\cite{ravikumar1996paralleldfa,
tewari2002paralleldfa}. Similar to Algorithm~\ref{alg:partref}, this algorithm
also performs partition refinement, but instead of doing so using leader
elections, it repeatedly computes a \emph{signature} for every state, and sorts
the states w.r.t.\ their signatures. This method allows splitting a block in
more than two subblocks with the downside that each iteration takes more than
constant parallel time. 

The algorithm from~\cite{tewari2002paralleldfa} uses hashing to construct and
compare signatures. Since sorting arrays is a very native operation on GPUs, we
follow~\cite{ravikumar1996paralleldfa} and use a sorting approach to construct
the new blocks.

Algorithm~\ref{alg:sort} presents this approach as \texttt{sortPR}. Again, an array \textit{block} is created (l.\ref{l:sort-block}).
In addition, an array \textit{state} is used for sorting the states (l.\ref{l:sort-state}). The signature of
a state consists of a list of block IDs, one for each $a \in \Sigma$: $\mathit{signature}[q][a]$ is equal to $q'$
iff $q \xrightarrow{a} q''$ and $\mathit{block}[q''] = q'$.

The array \textit{new\_block} is used to store the results of assigning new blocks to states (l.\ref{l:sort-newblock}).
Finally, the current number of blocks is stored at l.\ref{l:sort-numblocks} in \textit{num\_blocks}.

Next, at l.\ref{l:sort-for-initblockstate}--\ref{l:sort-initstate}, \textit{block} and \textit{state} are initialised.
The block consisting of all accepting states is given ID 0, while the other states are assigned to block $1$
(l.\ref{l:sort-initblock}). All the states are added to \textit{state} at l.\ref{l:sort-initstate}.

In the loop at l.\ref{l:sort-repeat}--\ref{l:sort-until}, the partition
refinement is performed until a fix-point has been reached, i.e., the number of
blocks has not increased (l.\ref{l:sort-until}). In each iteration of this loop,
the following is performed. First, in parallel, the signatures are updated
(l.\ref{l:sort-for-makesigs}--\ref{l:sort-makesigs}). After that, \textit{state}
is sorted in parallel, using \textit{signature} to compare states. The
comparison function is given at
l.\ref{l:sort-compare-begin}--\ref{l:sort-compare-end}. First, states are
compared based on the block they reside in. If they reside in the same block,
then the blocks they can reach via outgoing transitions are compared. Note
that the for loop starting in (l.\ref{l:sort-compare-loop}) is sequential and
requires iterating over the alphabet letters in a fixed order.

Once \textit{state} has been sorted, the parallel \emph{adjacent difference} is computed and stored in
\textit{new\_block}. The result of this is that $\mathit{new\_block}[0] = \mathit{state}[0]$ and for all
$0 < i < n$, $\mathit{new\_block}[i] = \mathit{are\_neq}(\mathit{state}[i], \mathit{state}[i-1])$, with \textit{are\_neq}
as defined at l.\ref{l:sort-areneq-begin}--\ref{l:sort-areneq-end}.
Once $\mathit{new\_block}[0]$ has been reset to $0$ (l.\ref{l:sort-set0}), \textit{new\_block} contains only $0$'s
and $1$'s, with each $1$ identifying the start of a new block. At l.\ref{l:sort-scan}, an \emph{inclusive scan} is
performed in parallel, resulting in \textit{new\_block} having been updated in such a way that for each $0 \leq i < n$,
$\mathit{new\_block}[i] = \sum_{0 \leq j \leq i} \mathit{new\_block}'[j]$, with \smash{$\mathit{new\_block}'$} referring
to \textit{new\_block} at the start of executing l.\ref{l:sort-scan}.

Now, for all $0 \leq i < n$, $\mathit{new\_block}[i]$ contains the new block of state $\mathit{state}[i]$.
At l.\ref{l:sort-for-updateblock}--\ref{l:sort-updateblock}, \textit{block} is updated in parallel to reflect this.
As the largest new block ID can be found at $\mathit{new\_block}[n-1]$, this location can be used to determine
the new number of blocks at l.\ref{l:sort-until}.

\begin{algorithm}
\begin{algorithmic}[1]
  \Require{A DFA $A = (Q, \Sigma, \delta, F, q_0)$, where $\lvert Q \rvert = n$ and $\lvert \Sigma \rvert = k$}
  \Ensure{The minimal quotient automaton represented in the array $block$}

  \State $\mathit{block} ::$ Array$[n]$ of type $\Nat$ \label{l:sort-block}
  \State $\mathit{state} ::$ Array$[n]$ of type $Q$ \label{l:sort-state}
  \State $\mathit{signature} :: $ Array$[n][k]$ of type $Q$ \label{l:sort-signature}
  \State $\mathit{new\_block} :: $ Array$[n]$ of type $\Nat$ \label{l:sort-newblock}
  \State $\mathit{num\_blocks} := 2$ \label{l:sort-numblocks}
    \ParDo{$q\in Q$} \label{l:sort-for-initblockstate}
      \State $\mathit{block}[q] := (q \in F\ ?\ 0 : 1)$ \Comment Initialise \label{l:sort-initblock}
      \State $\mathit{state}[q] := q$ \label{l:sort-initstate}
    \EndParDo
  \Repeat \label{l:sort-repeat}
    \State $\mathit{num\_blocks} := \mathit{new\_block}[n-1] + 1$ \Comment Number of blocks before iteration
    \ParDo{$(q, a) \in Q \times \Sigma$} \label{l:sort-for-makesigs}
        \State $\mathit{signature}[q][a] := block[\delta(q, a)]$ \label{l:sort-makesigs}
    \EndParDo
      \State $\mathtt{sort}(\mathit{state}, \textsc{compare})$ \label{l:sort-sortstates}
      \State $\mathit{new\_block} := \mathtt{adjacent\_diff}(\mathit{state}, \textsc{are\_neq})$ \Comment Place $1$ for each change \label{l:sort-diff}
      \State $\mathit{new\_block}[0] := 0$ \label{l:sort-set0}
      \State $\mathit{new\_block} := \mathtt{inclusive\_scan}(\mathit{new\_block})$ \Comment Compute new block labels \label{l:sort-scan}
      \ParDo{$q\in Q$} \label{l:sort-for-updateblock}
        \State $\textit{block}[state[q]] = \textit{new\_block}[q]$ \label{l:sort-updateblock}
      \EndParDo
  \Until{$\mathit{new\_block[n-1]} + 1 = \mathit{num\_blocks}$} \label{l:sort-until}
\vspace{0.4cm}
\Function{compare}{$q_1$, $q_2$} \label{l:sort-compare-begin}
\If{$\mathit{block}[q_1] > \mathit{block}[q_2]$}
	\Return $\mfalse$
\EndIf
\If{$\mathit{block}[q_1] < \mathit{block}[q_2]$}
	\Return $\mtrue$
\EndIf
\ForAll{$a \in \Sigma$} \label{l:sort-compare-loop}
	\If{$\mathit{signature}[q_1][a] > \mathit{signature}[q_2][a]$}
		\Return $\mfalse$
	\EndIf
	\If{$\mathit{signature}[q_1][a] < \mathit{signature}[q_2][a]$}
		\Return $\mtrue$
	\EndIf
\EndFor
\State \Return $\mfalse$
\EndFunction \label{l:sort-compare-end}
\vspace{0.4cm}
\Function{are\_neq}{$q_1$, $q_2$} \label{l:sort-areneq-begin}
\If{$\mathit{block}[q_1] \neq \mathit{block}[q_2]$}
	\Return $\mtrue$
\EndIf
\ForAll{$a \in \Sigma$}
	\If{$\mathit{signature}[q_1][a] \neq \mathit{signature}[q_2][a]$}
		\Return $\mtrue$
	\EndIf
\EndFor
\State \Return $\mfalse$
\EndFunction \label{l:sort-areneq-end}
\end{algorithmic}
\caption{\label{alg:sort} Parallel sorting-based algorithm~$\mathtt{sortPR}$}
\end{algorithm}

\noindent
In~\cite{ravikumar1996paralleldfa} it is shown that on average this algorithm
has polylogarithmic run-time complexity. The argument given uses the fact that
on uniformly sampled DFAs almost all pairs of states have a shortest
distinguishing word of polylogarithmic depth. This fact is attributed
to~\cite{trachtenbrot1973finite}. Although this is true for uniformly sampled
DFAs, we like to stress that for many use cases and real-life applications this
bound does not apply. For example, in the Fibonacci automata presented in
Section~\ref{sec:results} this is not the case. In the automaton $\Fib_i$
containing $n$ states, there is a pair of states for which the shortest
distinguishing word, and thus also the number of iterations, has length $n{-}2$.

\subsection{Partition refinement using partial transitive closure}
In this section, we present a new DFA minimisation algorithm \texttt{transPR}. The main idea of
the algorithm is to perform partition refinement like the algorithms before, 
but in the initialisation compute the transitive closure on each distinct
 letter. After this initialisation step, we use $\texttt{naivePR}$ to
 complete the minimisation.
  
 This approach is presented in Algorithm~\ref{alg:partref-trans}. This
 is done in a data-parallel way which is also used for prefix sum and finding
 the end of a linked list~\cite{Hillis86}. On some DFAs, with a rather specific
 structure, this method exponentially improves the runtime compared to the
 other partition refinement algorithms.

\begin{algorithm}
  \begin{algorithmic}[1]
    \State $\Sigma^T := \{a^{2^i} \mid a\in \Sigma, i \in [0, \lfloor\log n \rfloor]\}$
    \State $\delta^T :: Q \times \Sigma^T \mapsto Q$ 
    \State $\delta^T(q, a) := \delta(q,a)$ for all $a\in \Sigma$
    \ForAll{$i \in [1, \lfloor\log n \rfloor]$}
      \ForAll{$a\in \Sigma$}
        \ParDo{$q \in Q$}
          \State $\delta^T(q, a^{2^i}) := \delta^T(\delta^T(q, a^{2^{i-1}}),a^{2^{i-1}})$\label{l:log-trans-closure}
        \EndParDo
      \EndFor
    \EndFor
    \State Perform $\texttt{naivePR}$ on the DFA $A' = (Q, \Sigma^T, \delta^T, F, q_0)$
  \end{algorithmic}
  \caption{\label{alg:partref-trans} Parallel partition refinement with transitive closure~$\mathtt{transPR}$}
  \end{algorithm}

\noindent
The algorithm works by adding letters for increasingly large words of the same
letters. Given an input DFA $A= (Q, \Sigma, \delta, F, q_0)$, we construct a DFA
$A' = (Q, \Sigma^T, \delta^T, F, q_0)$ which has the same set of states and
final states, but a larger alphabet $\Sigma^T$. The alphabet contains the
letters $a^{2^0}, a^{2^1}, \dots, a^{2^{\lfloor \log n\rfloor}}$ for each
original letter $a\in \Sigma$. The transition function is computed such that for
each new symbol $a^k\in \Sigma^T$ the transition function $\delta^T(q_1,a^k) =
q_k$ if in the original DFA $Q$ there are states  $q_1, \dots q_k \in Q$ such
that $\delta(q_{i}, a) = q_{i+1}$ for each $i\in[1,k]$. This can be computed in
a logarithmic number of parallel steps, by using the previously computed
transitions, as is done at l.\ref{l:log-trans-closure} of
Algorithm~\ref{alg:partref-trans}.

The correctness of this algorithm relies on the fact that equality on states is
invariant under the partial closure that is added. Indeed, we can see that the
DFA $A'$ obtained in Algorithm~\ref{alg:partref-trans} is language equivalent
to the input DFA $A$ if we consider the alphabet letters added as words. If
$\delta^T(q,a_T) = q'$ for some $a_T \in \Sigma^T$, then $a_T= a^{2^j}$ for some
$a\in \Sigma$ and $j \in [0, \lfloor \log n\rfloor ]$.
By construction there is a sequence $q_0, \dots q_{k}$ such that $q_0 = q$,
$q_{i+1} = \delta(q_i,a)$ and $q_{k} = q'$. 

\begin{figure}
  \resizebox{\textwidth}{!}{
\begin{tikzpicture}[node distance= 1.5cm, initial text = ]
  \node[state,initial] (0) at (0,0) {$q_0$};
  \node[state] (1) [right of=0]{$q_1$};
  \node[state] (2) [right of=1]{$q_2$};
  \node[state] (3) [right of=2]{$q_3$};
  \node[state] (4) [right of=3]{$q_4$};
  \node[state] (5) [right of=4]{$q_5$};
  \node[state] (6) [right of=5]{$q_6$};
  \node[state] (7) [right of=6]{$q_7$};
  \node[state] (8) [right of =7] {$q_8$};
  \node[state, accepting] (9) [right of =8] {$q_9$};

  \path[->]
  (0) edge node [above]{$a$} (1) 
  (1) edge node [above]{$a$} (2) 
  (2) edge node [above]{$a$} (3) 
  (3) edge node [above]{$a$} (4) 
  (4) edge node [above]{$a$} (5) 
  (5) edge node [above]{$a$} (6) 
  (6) edge node [above]{$a$} (7) 
  (7) edge node [above]{$a$} (8) 
  (8) edge node [above]{$a$} (9) 
  (9) edge[loop above] node [above]{$a$} (9);

  \path[->, dashed] 
  (0) edge [bend right=30] node[below] {$a^2$} (2) (0) edge [bend left]
  node[above] {$a^4$} (4) (0) edge [bend right=40] node[below] {$a^8$} (8);
  \end{tikzpicture}} \caption{The DFA $A=(\{q_0, \dots , q_9\}, \{a\}, \delta,
  \{q_9\}, q_0)$ with the extra partial transitive closure from $q_0$ added in dashed
  arrows.\label{fig:dfalong}}
\end{figure}
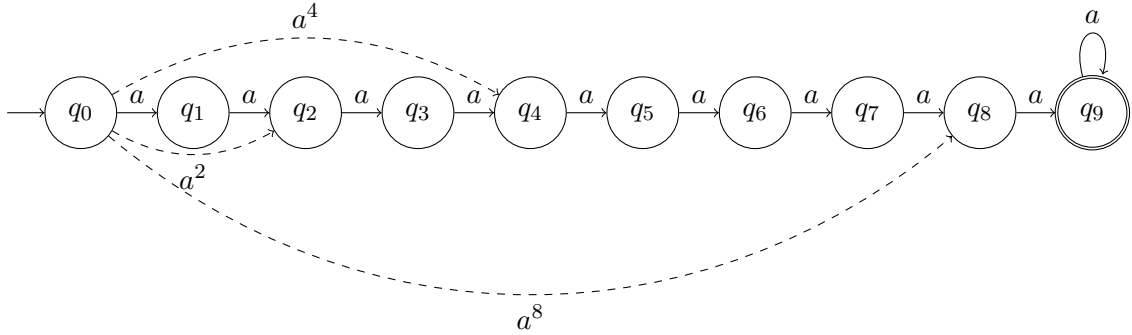

This approach helps in the case of long paths with the same letter. Consider the
DFA $A$ from Figure~\ref{fig:dfalong}. This DFA accepts all words $a^j$ with
$j > 8$. Any parallel partition refinement algorithm would need at least eight
iterations to conclude that $q_0$ is not the same as $q_1$.  However, building
the partial transitive closure only requires a logarithmic number of parallel
iterations. With this partial transitive closure added, a partition refinement
algorithm can in the first iteration conclude that $q_0$ is different from $q_1$
since the transition with $a^8$ leads to different states.  

\section{DFA equivalence checking and inclusion checking}
\label{sec:dfa-eq}

So far, we have focussed on DFA minimisation algorithms. Another classical operation on DFAs
is \emph{equivalence checking}, which has its practical uses in formal verification, in particular
model checking, and compiler construction.

DFA equivalence checking entails checking whether or not two given DFAs have the same
language. This can be done via DFA minimisation~\cite{hopcroft1971DFAmin}, but it can also
be performed with the well-known Hopcroft-Karp algorithm~\cite{hopcroft-karp,AhHoUl74}.
Algorithm~\ref{alg:hopcroft-karp} presents this algorithm. Given two DFAs $A$ and $B$,
a relation $R$ is built such that it relates states of $A$ with states of $B$. Specifically, the algorithm
tries to build a bisimulation relation between the DFAs. If it succeeds in doing so, the DFAs
accept the same language.

At l.1, $R$ is initialised as an empty set, and another set $S$ is created to store the pairs
of states that still need to be processed. In the while-loop at lines 2--9, this processing of pairs
in $S$ is performed. At l.3, an arbitrary pair $\langle q^A, q^B \rangle$ is selected and
removed from $S$. If the two states $q^A$ and $q^B$ are already related by $R^{+}$, i.e., 
the transitive closure of $R$, then no work needs to be done for this pair.
Otherwise, it must be checked whether or not exactly one of the two states is accepting
(l.5). If that is the case, then $A$ and $B$ do not have the same language, and $\mfalse$
is returned. Otherwise, for all $a \in \Sigma$, the states $\delta^A(q^A, a)$ and $\delta^B(q^B, a)$
are combined in a new pair and added to $S$ (lines 7--8). By doing so, checks are added
so that, if a particular letter is followed from both $q^A$ and $q^B$, then the states reached by doing
so are again bisimilar. Next, $\langle q^A, q^B \rangle$ is added to $R$.
If all the checks in $S$ succeed, eventually $S$ will be empty, and hence
$\mtrue$ will be returned (l.10).

\begin{algorithm}
  \begin{algorithmic}[1]
  \Require{DFAs $A = (Q^A, \Sigma^A, \delta^A, F^A, q_0^A)$ and $B = (Q^B, \Sigma^B, \delta^B, F^B, q_0^B)$}
  \Ensure{$\mtrue$ iff $\Lang(A) = \Lang(B)$}
    \State $R := \emptyset; S := \{ \langle q_0^A, q_0^B \rangle \}$
    \While{$S \neq \emptyset$}
    	\State Select $\langle q^A, q^B \rangle$ from $S$; $S := S \setminus \{ \langle q^A, q^B \rangle \}$
	\If{$\langle q^A, q^B \rangle \not\in R^{+}$}
		\If{$(q^A \in F^A \wedge q^B \not\in F^B) \vee (q^A \not\in F^A \wedge q^B \in F^B)$}
			\Return $\mfalse$
		\Else
			\ForAll{$a \in \Sigma$}
				\State $S := S \cup \{ \langle \delta^A(q^A, a), \delta^B(q^B, a) \rangle \}$
			\EndFor
			\State $R := R \cup \{ \langle q^A, q^B \rangle \}$
		\EndIf
	\EndIf
    \EndWhile
    \State \Return $\mtrue$
  \end{algorithmic}
  \caption{\label{alg:hopcroft-karp} Hopcroft-Karp DFA equivalence checking algorithm}
  \end{algorithm}

\noindent
The check at l.4 requires maintaining the transitive closure of $R$. Hopcroft and
Karp achieve this with a union-find data structure, to keep track of a set of state equivalence
classes instead of a set of individual state pairs. As state pairs are added to $R$ at l.9,
these equivalence classes are updated. The use of this data structure results in the algorithm
having a near-linear complexity.

Maintaining such a union-find data structure in a massively parallel way is highly non-trivial.
An alternative is to maintain the set of state pairs that are added to $R$ without updating its
transitive closure.
While this negatively impacts the
algorithm's complexity, this may be mitigated by performing the algorithm massively parallel
on a GPU. In practice, sets of elements are typically maintained by means of a hash table.
Although not commonly used, hash tables for GPUs have been developed in recent
years~\cite{lessley-survey,awad_analyzing_2023,WijOs23,europar24}. For this article, we use a GPU hash table developed for the explicit-state
model checker \GPUexplore~\cite{WijOs23,gpuexplore3,gpuexplore3-journal}. This model checker accepts as input models that consist
of a finite number of state machines that can synchronise on transition actions and that can
read from and write to Integer and Boolean variables. Next, we show how the problem of
checking whether two DFAs are equivalent via the `naive' version of the
Hopcroft-Karp algorithm, i.e., the version in which a set of state pairs is maintained, can be
encoded as an input model for \GPUexplore. Given such a model, \GPUexplore will construct
its state space, which corresponds directly to the construction of the relation $R$. An on-the-fly
check corresponding to l.5 of Algorithm~\ref{alg:hopcroft-karp} achieves that
\GPUexplore terminates early and reports a problem whenever the two DFAs are not equivalent.

\begin{figure}
  \begin{center}
    \resizebox{0.30\textwidth}{!}{
  \begin{tikzpicture}[node distance= 3cm, initial text = ]
    \node[state,initial] (00) at (0,0) {$00$};
    \node[state] (01) [right of=00]{$01$};
    \node[state, accepting] (11) [below of=00]{$11$};
    \node[state, accepting] (10) [below of=01]{$10$};

    \path[->]
    (00) edge node [above]{$1$} (01)
    (00) edge [loop above] node {$0$} (00)
    (01) edge node [above left, pos=0.2]{$1$} (11)
    (01) edge node [above left ]{$0$} (10)
    (10) edge node [above right, pos=0.2]{$0$} (00)
    (10) edge [bend right]  node [left] {$1$} (01)
    (11) edge node [above]{$0$} (10)
    (11) edge [loop above] node {$1$} (11);
    \end{tikzpicture}}
    \end{center}
    \caption{The DFA $\memory_2$.\label{fig:mem2}} 
\end{figure}
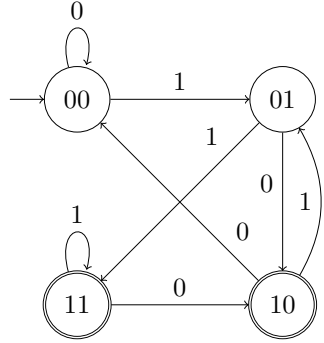

\begin{figure}
\begin{lstlisting}[language=slco]
model memory_2 {
  actions t, f
  classes
    P {
      variables
        Boolean b0_0, b1_0, b0_1, b1_1
      state machines
        SM0 {
          forbidden Error initial S
          transitions
            S -> S { <f> [b1_0 := b0_0; b0_0 := false] }
            S -> S { <t> [b1_0 := b0_0; b0_0 := true] }
            S -> Error { (b1_0 and not b1_1) or (not b1_0 and b1_1) }
        }
  	    SM1 {
  	      initial S
          transitions
            S -> S { <f> [b1_1 := b0_1; b0_1 := false] }
            S -> S { <t> [b1_1 := b0_1; b0_1 := true] }
  	    }
    }
  objects p: P()
}
\end{lstlisting}
\caption{An \SLCO model to check equivalence of two instances of $\memory_2$.}
\label{fig:mem2_slco}
\end{figure}

\noindent
A variant of the algorithm by Hopcroft and Karp can be used to perform \emph{inclusion checking},
i.e., checking that the language of the first DFA is included in the language of the second DFA.
To do this, the condition at l.5 of Algorithm~\ref{alg:hopcroft-karp} must be changed to
$a^A \in F^A \wedge q^B \not\in F^B$. In other words, the second disjunct of the original condition
must be dropped.

An example DFA is given in Figure~\ref{fig:mem2}. This DFA is instance 2 of the $\memory$
benchmark, i.e., $\memory_2$ (see Section~\ref{sec:benchmarks}). It reads a sequence of bits, and remembers
the last two bits it has read. In the initial state, those two bits are initialised to $0$.

Figure~\ref{fig:mem2_slco} presents a way to encode equivalence checking of two instances
of $\memory_2$.
This is done using the \emph{Simple Language of Communicating Objects} (\SLCO)~\cite{engelenthesis,facs18,facs23,slco26}, which is
one of the accepted input languages of \GPUexplore.
A model may have actions (see l.2), in this case \slcoinline!t!, corresponding
to $1$, and \slcoinline!f!, corresponding to $0$.
Furthermore, a model can contain a number of \emph{classes}. In this case, there is only
the class \slcoinline!P!. A class, in turn, may have a number of variables. For this example,
variables are needed to store the two most recent bits, for both instances of $\memory_2$,
therefore the Boolean variables \slcoinline!b0_0!, \slcoinline!b1_0!, \slcoinline!b0_1! and
\slcoinline!b1_1! are declared at l.6. Two state machines, representing the two DFAs,
are given at lines 8--14 and lines 15--20.

For the first state machine, \slcoinline!SM0!, an \slcoinline!initial! state \slcoinline!S! and a \slcoinline!forbidden! state \slcoinline!Error! are declared at l.9. Two transitions represent the
transitions of the DFA, at lines 11--12. The notation
\slcoinline!S -> S { <a> [x_0 := E_0; x_1 := E_1] }! should be interpreted as follows.
If the state machine is in state \slcoinline!S!, then it can transition to state \slcoinline!S! if
action \slcoinline!a! can be fired, and when this occurs, the assignments
\slcoinline!x_0 := E_0! and \slcoinline!x_1 := E_1! will be executed in order, in one atomic step.

Note at l.13 the transition to the \slcoinline!forbidden! state \slcoinline!Error!.
This transition is enabled iff the expression associated with it, between the curly brackets,
evaluates to $\mtrue$. Note furthermore that this condition corresponds to the condition at l.5
of Algorithm~\ref{alg:hopcroft-karp}. Whenever the \slcoinline!Error! state is reached when
performing state space exploration, \GPUexplore will terminate and report the error, which in
this case represents the in-equivalence of the two DFAs.

State machine \slcoinline!SM1! is almost a copy of \slcoinline!SM0!, apart from the use of an
\slcoinline!Error! state. Since \slcoinline!SM1! has the same actions \slcoinline!t! and
\slcoinline!f! associated with transitions, and therefore in its alphabet, \slcoinline!t! and
\slcoinline!f! need to synchronise: a \slcoinline!t!-transition (\slcoinline!f!-transition) of
\slcoinline!SM0! can be followed iff a \slcoinline!t!-transition (\slcoinline!f!-transition) of
\slcoinline!SM1! can be followed, and when this is done, the two state machines transition
together in one step. This mechanism achieves that exploration of the state space of
model \slcoinline!memory_2! corresponds to constructing the synchronous product
of \slcoinline!SM0! and \slcoinline!SM1!, and therefore of the two DFAs.

Finally, at l.22, an instance of class \slcoinline!P!, the \slcoinline!object p!, is created.
In \SLCO, classes can be instantiated multiple times, but in this application, only a single
class instance is necessary.  

Note in the example that the DFAs, as described in an \SLCO model, are not explicitly given,
but instead described at a higher level, i.e., specified. In this particular example, for instance, the four states of DFA
$\memory_2$ are not explicitly represented by four states of an \SLCO state machine,
and its eight transitions are also not represented by eight \SLCO transitions directly,
but instead captured by the two transitions
at lines 11--12 (and 18--19) of Figure~\ref{fig:mem2_slco}. While this means that in our
approach state space exploration does not only entail computation of the synchronous
product, but also the construction of the actual input DFAs, we argue that this is not only
practically necessary, but also realistic. An explicit encoding of a DFA in \SLCO is feasible,
but as the DFAs become larger, such an encoding would quickly become unwieldy, and
\GPUexplore would not be able to process the resulting models. In practice, however, for
the vast majority of DFAs, a high-level description is available from which a specification
can be derived as we have in our example \SLCO model. In other words, typically, a DFA
is not a randomly constructed automaton, but instead there is some rational behind its creation,
which can be used to construct a specification.

To perform inclusion checking with \GPUexplore, the condition to transition
to the \slcoinline!Error! state must be changed similarly to the condition at l.5
of Algorithm~\ref{alg:hopcroft-karp}. For instance, the transition at l.13 of
Figure~\ref{fig:mem2_slco} should become \slcoinline!S -> Error { b1_0 and not b1_1 }!.

\begin{figure}[t]
\centering
\includegraphics[width=\textwidth]{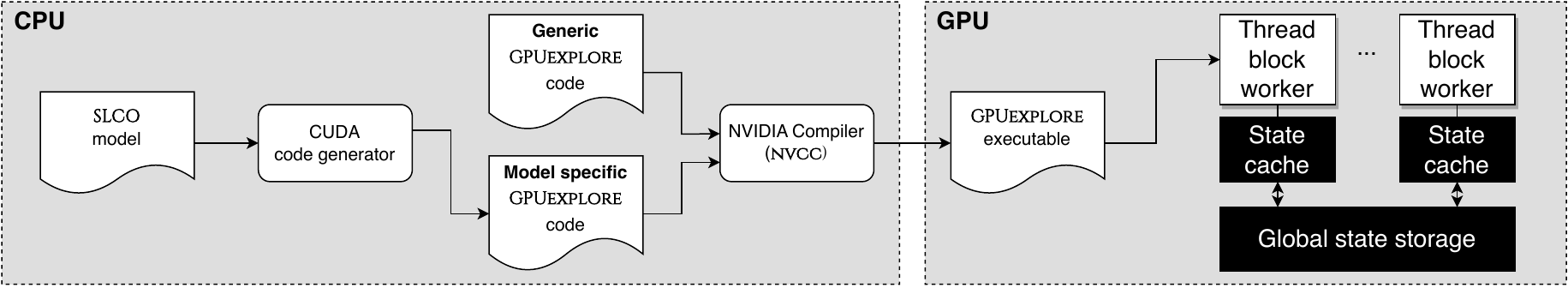}
\caption{The workflow from \SLCO model to \GPUexplore model checking.}
\label{fig:gpuexplore}
\end{figure}

\paragraph{\GPUexplore.}
Figure~\ref{fig:gpuexplore} presents the workflow from \SLCO model to model checking
with \GPUexplore. Given an \SLCO model, a code generator, written in \textsc{Python},
generates CUDA \CC code to interpret the model: for every transition in the model, code is
generated to check whether the transition can be fired, and to actually fire the transition.
This code is combined with generic code, i.e., code not specific for a particular \SLCO model.
This code contains, for instance, an implementation of the used GPU hash table.
Next, NVIDIA's NVCC compiler compiles the combined code, resulting in an executable that
can be launched on an NVIDIA GPU. In practice, it has been shown that GPU-acceleration
of explicit-state model checking can be very efficient, with speedups of hundreds of times
compared to state-of-the-art CPU-based explicit-state model checkers~\cite{gpuexplore3,gpuexplore3-journal,gpuexplore-ltl}.

On the right in Figure~\ref{fig:gpuexplore}, an overview is given of how \GPUexplore runs
on a GPU. A global hash table, containing the constructed states, resides in global memory.
Many thread blocks, each containing 512 threads, scan this hash table for states that require
exploration, i.e., for which their outgoing transitions need to be checked and potential
successor states need to be constructed. States that require exploration are temporarily stored
in \emph{shared memory}, which is fast, but small, on-chip memory that the threads in a block
can use together. Any constructed successor states are stored in a shared memory hash table,
before they are synchronised with the global memory hash table. This process is massively parallel, with the threads in each block working together on both successor generation and global memory storage. This is repeated
until all the states reachable from the initial state have been identified.
More information on \GPUexplore can be found in~\cite{GPUExplore,GPUExplore2,GPUexplore-safety,gpuexplore3,WijOs23,gpuexplore-ltl,gpuexplore3-journal}.

\section{Experiments}
\label{sec:exps}

In this section, we discuss the results of our implementations.\footnote{The source code of the implementations, including the DFA families used for benchmarking,
are available at \url{https://gitlab.tue.nl/awijs/dfamin.git}.} We benchmarked
the implementations of DFA minimisation algorithms with respect to three families of DFAs: Fibonacci DFAs $\Fib_n$
from~\cite{castiglione2008hopcroft}, bit-splitters $\mathcal{B}_n$ derived
from~\cite{groote2023lowerbounds}, and DFAs derived from a subset of the VLTS
benchmark set.\footnote{\url{https://cadp.inria.fr/resources/vlts} (visited on: 07-2026).} Our approach to DFA equivalence checking has been benchmarked using
two families of DFAs:  the \emph{Cycle DFAs} $\C_n$, and an extension of the bit-splitters $\mathcal{B}_n'$.
The $\C_n$ family is inspired by the Fibonacci DFAs,
but contrary to the latter, these DFAs can have more than one letter and do not involve Fibonacci words over the binary alphabet.
The $\mathcal{B}_n'$ family extends the bit-splitters with an
initial state, as the original bit-splitters do not have one.

For inclusion checking, we used the family of DFAs $\memory_n$.
All these families are explained in detail in Section~\ref{sec:benchmarks}.

The reason that, except for the VLTS benchmarks, all of the families of DFAs that we use are artificially constructed, instead of representing
`real' systems,
is a current lack of public benchmark suites consisting of large DFAs, i.e., in the order of millions of states. Furthermore, in each
equivalence check, we compare a DFA with itself, meaning that the explored state space is exactly as large as the DFA itself.
While this only results in a linear growth of the state space as the DFA grows in size, the benefit of this setup is that the equivalence
checking algorithm will not terminate early, as the two DFAs are equivalent, which allows for better scalability measurements.
Furthermore, as our DFAs are encoded symbolically, i.e.,
using variables, as in Figure~\ref{fig:mem2_slco}, the actual state space still needs to be constructed, it cannot be directly derived from
the input DFAs. Finally, as \GPUexplore can handle large state spaces, in the order of hundreds of millions to billions of states,
very well~\cite{gpuexplore3,gpuexplore3-journal}, equivalence
checking is expected to scale in a very similar way.

\subsection{Benchmarks}
\label{sec:benchmarks}

\paragraph{Fibonacci DFAs:} The first family of DFAs we use for benchmarking
consists of so-called Fibonacci automata. These are simple automata with only a
unary alphabet. However, they exhibit very particular behaviour. As witnessed
in~\cite{castiglione2008hopcroft}, these automata are notoriously hard for
partition refinement and the number of iterations of any partition refinement
algorithm is $n$. The automata are called Fibonacci automata due to the close
correspondence to Fibonacci words over the binary alphabet, which are defined
inductively as follows: the base cases are $w_0 = 1$, and $w_1 = 0$, and for every $i\in \Nat \setminus \{ 0 \}$,
$w_{i+1} = w_iw_{i-1}$. This gives the following sequence:
\begin{align*}
  w_2 &= 01 \\
  w_3 &= 010 \\ 
  w_4 &= 01001 \\
  w_5 &= 01001010 \\
  \ldots
\end{align*}

\noindent
For every $n\in \Nat$, we define the automaton $\Fib_n = (Q, \{a\}, \delta, q_0, F)$ as follows,
with $w_n[i]$ referring to the $i$-th bit in the bit sequence $w_n$:
\begin{itemize}
  \item the set of states is $Q =\{q_i \mid i \in [0,|w_n|-1 ] \}$; 
  \item the transition function is $\delta(q_i,a) = q_{i+1\mod |w_n|}$;
  
  \item the set of final states is $F = \{q_i \mid q_i \in Q\text{ and } w_n[i] = 1\}$.
\end{itemize}
\newcommand{\bit}[1]{\mathtt{#1}}

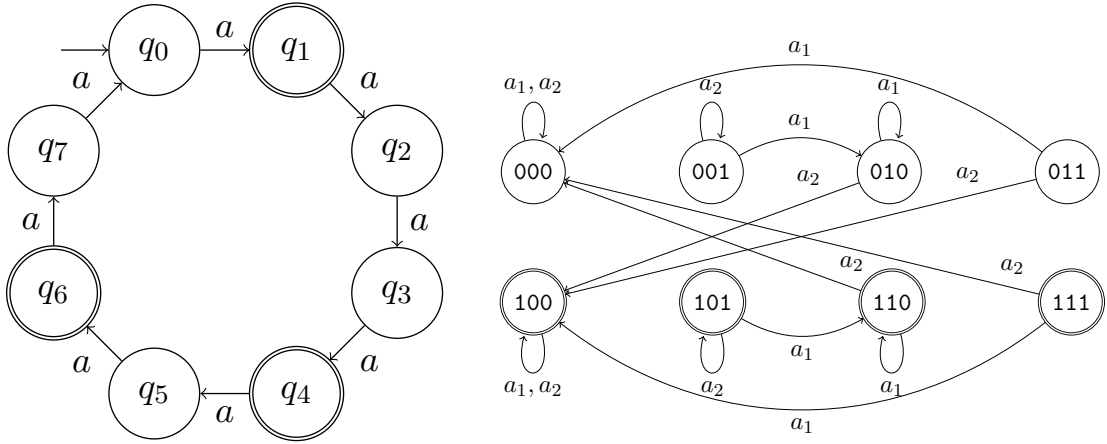
\begin{figure}
  \begin{center}
    \resizebox{0.39\textwidth}{!}{
  \begin{tikzpicture}[node distance= 1.5cm, initial text = ]
    \node[state,initial] (0) at (0,0) {$q_0$};
    \node[state, accepting] (1) [right of=0]{$q_1$};
    \node[state] (2) [below right of=1]{$q_2$};
    \node[state] (3) [below of=2]{$q_3$};
    \node[state,,accepting] (4) [below left of=3]{$q_4$};
    \node[state] (5) [left of=4]{$q_5$};
    \node[state,accepting] (6) [above left of=5]{$q_6$};
    \node[state] (7) [above of=6]{$q_7$};

    \path[->]
    (0) edge node [above]{$a$} (1) 
    (1) edge node [above right ]{$a$} (2)
    (2) edge node [right]{$a$} (3)
    (3) edge node [below right]{$a$} (4) 
    (4) edge node [below]{$a$} (5)
    (5) edge node [below left]{$a$} (6)
    (6) edge node [left]{$a$} (7)
    (7) edge node [above left]{$a$} (0);
    \end{tikzpicture}}
\resizebox{0.6\textwidth}{!}{
    \begin{tikzpicture}[node distance=1.75cm] 

      \node[state] (000) {$\bit{000}$};
    
      \node[state, right=of 000]  (001) {$\bit{001}$};
      \node[state, right= of 001] (010) {$\bit{010}$};
      \node[state, right= of 010] (011) {$\bit{011}$};
      \node[state, accepting, below= 1cm of 000]  (100) {$\bit{100}$};
      \node[state, accepting, right=of 100]  (101) {$\bit{101}$};
      \node[state, accepting, right=of 101]  (110) {$\bit{110}$};
      \node[state, accepting, right=of 110]  (111) {$\bit{111}$};
    
      \node[below left =0.5cm and 0.5cm of 000]  (lineleft) {};
      \node[above right=0.5cm and 0.5cm of 111] (lineright) {};
        
      \path[->] 
      (000) edge [loop above] node {$a_1, a_2$} (000)
      
      (001) edge [loop above] node {$a_2$} (001)
      (001) edge [bend left]  node [above] {$a_1$} (010)
      
      (010) edge node [pos=0.1, above left] {$a_2$} (100)
      (010) edge [loop above] node [above] {$a_1$} (010)
      
      (011) edge node [pos=0.1, above left]{$a_2$} (100)
      (011) edge [bend right=38] node[above] {$a_1$} (000)
      
      (100) edge [loop below] node{$a_1, a_2$} (100)
      
      (101) edge [bend right] node[below] {$a_1$} (110)
      (101) edge [loop below] node[below] {$a_2$} (101)
      
      (110) edge node[pos=0.1, above right] {$a_2$} (000)
      (110) edge [loop below] node [below] {$a_1$} (110)
      
      (111) edge node[pos=0.1, above right] {$a_2$} (000)
      (111) edge[bend left=38] node[below] {$a_1$} (100)
       ;
    \end{tikzpicture}}
    \end{center}
    \caption{The DFA $\Fib_5$ on the left, and the DFA $\mathcal{B}_3$ on the
    right.\label{fig:dfas}} 
\end{figure}

\noindent
The Fibonacci DFA $\Fib_5$ is given in Figure~\ref{fig:dfas}.

\paragraph{Bit-splitters:} 
The second family of automata consists of the so-called \emph{bit-splitters}
$\mathcal{B}_n$. For $n\in \Nat$, the bit-splitter $\mathcal{B}_n$ is a
deterministic automaton with $2^n$ states and an alphabet consisting of $n{-}1$
symbols. By construction, during partition refinement, every time a block can be
split, it is split in two blocks of equal size. This property makes the family
inherently hard for partition refinement algorithms. However, the parallel
algorithm requires only a logarithmic number of iterations to compute the
minimal DFA. The bit splitter $\mathcal{B}_3$ is given in Figure~\ref{fig:dfas}.

The family does not contain an initial state, and comes from the setting of
Labelled Transition Systems (LTSs).
An LTS is a graph structure with a finite number of states and transitions
between states, with each transition having an action label.

Since it is a hard example for partition refinement, we use it for this
purpose and allow the absence of an initial state.
In the following, states $\sigma$ represent bit sequences of length $n$,
i.e., $\sigma \in \{0, 1\}^n$.
We define $\mathcal{B}_1= (Q_1,
\Sigma_1, \delta_1, F_1)$, where $Q_1 = \{\bit0, \bit1\}$, $\Sigma_1 =
\emptyset$ and $F_1=\{\bit1\}$. Given the automaton $\mathcal{B}_n = (Q_n, \Sigma_n,
\delta_n, F_n)$ for some $n \in \Nat$,
we define $\mathcal{B}_{n+1}=(Q_{n+1},
\Sigma_{n+1}, \delta_{n+1}, F_{n+1})$, such that:
\begin{itemize}
  \item The set of states contains two copies of $Q_n$, i.e., $Q_{n+1} = \{
  \bit0\sigma, \bit1\sigma \mid \sigma\in Q_n \}$,
  \item One fresh symbol $a_n \not\in \Sigma_n$ is added to the alphabet: $\Sigma_{n+1} = \Sigma_{n}
  \cup \{ a_n \}$,
  \item The transition function $\delta_{n+1}$ is defined such that for each
  $a_m \in \Sigma_n$, and state $\bit{b}\sigma \in Q_{n+1}$, it maintains the
  behaviour of $\mathcal{B}_{n}$, i.e., $\delta_{n+1}(\bit{b}\sigma, a_m) =
  \bit{b}\delta_n(\sigma, a_m)$. For the fresh symbol $a_n \in \Sigma_{n+1}
  \setminus \Sigma_n$, $\delta_{n{+}1}$ is extended as follows, with $\bit{\bar{b}}$ being
  the bit flipped version of $\bit{b}$, i.e., $\bit{b} = 0 \iff \bit{\bar{b}} = 1$:
  \[
    \delta_{n{+}1}(\bit{b}\sigma, a_n)= \begin{cases}
      \bit{\bar{b}} \bit0^n & \text{If } \sigma[0] = 1,\\
      \bit{b}\sigma & \text{otherwise.}
    \end{cases}
  \]
  \item the set of accepting states is $F_{n+1} = \{\bit1 \sigma \mid \sigma \in
  Q_n \}$.
\end{itemize}

\noindent
As previously mentioned, bit-splitter DFAs are constructed in such a way that they are inherently hard to
minimise by partition refinement. Each bit-splitter $\mathcal{B}_{n+1}$ combines
two copies of the bit-splitter $\mathcal{B}_n$, with the transition
function defined in such a way that each possible split divides an existing block in two
blocks of the same size. This results in a DFA in which the amount of required work for splitting is large,
since each split involves moving many states to the new block. However, because in each split a block is split in two parts of the same size, the number of sequential splits
needed is smaller than for the Fibonacci automata.

\paragraph{VLTSs:}
For the benchmarking of our DFA minimisation implementations, we also use the VLTS benchmark
suite. The VLTS acronym stands
for \emph{Very Large Transition Systems}. This suite consists of LTSs that
originate from modelling protocols and concurrent systems. Some of the
benchmarks are from case studies from industrial systems. 

The transition
relation of an LTS does not need to be deterministic, nor complete. We turn an LTS into a
DFA by first making the LTS deterministic such that each state has at most one
outgoing transition for every label, using the powerset construction
algorithm~\cite{powerset}. To convert the deterministic LTS to a DFA, we need to
complete the transition function and define which states are accepting. We
define all the states as accepting and add one new non-accepting state $\bot$.
For each state $q$ and label $a$ for which there exists no transition with that
label from $q$, we add a new transition labelled $a$ to $\bot$, i.e.
$\delta(q,a) = \bot$. This completes the transition function and creates a DFA
accepting all the words corresponding to a path through the original LTS. 

Due to state-space explosion we were not able to make all VLTS benchmarks
deterministic. We used all benchmarks for which the computation to make them
deterministic took less than ten minutes.

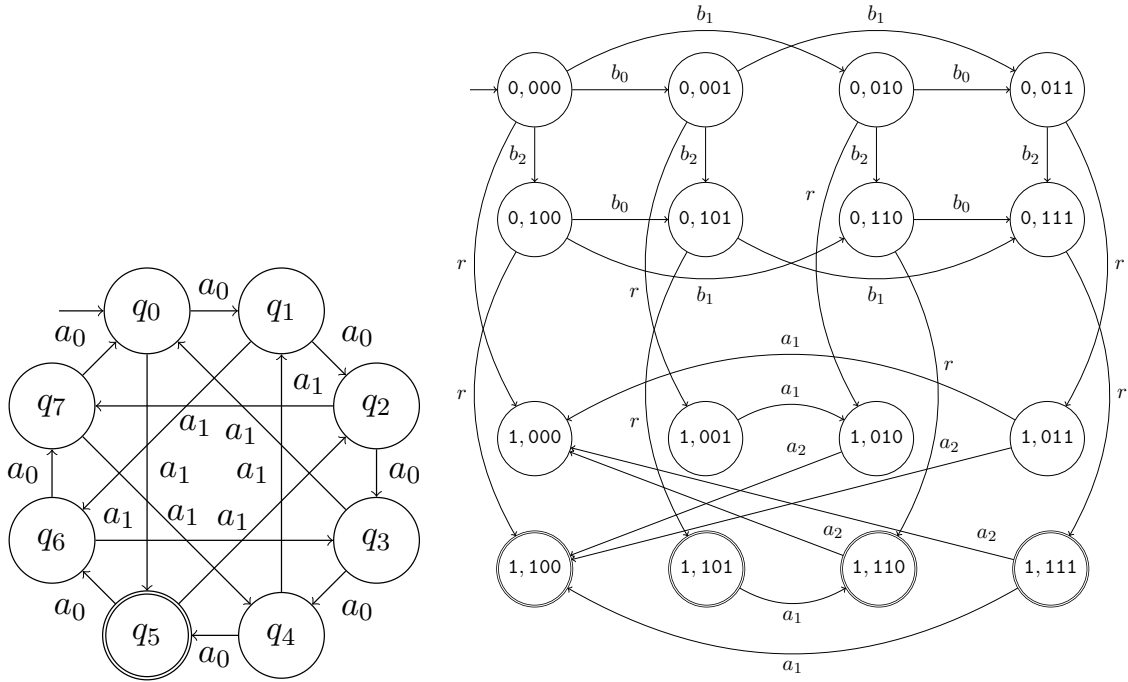
\begin{figure}
  \begin{center}
    \resizebox{0.39\textwidth}{!}{
  \begin{tikzpicture}[node distance= 1.5cm, initial text = ]
    \node[state,initial] (0) at (0,0) {$q_0$};
    \node[state] (1) [right of=0]{$q_1$};
    \node[state] (2) [below right of=1]{$q_2$};
    \node[state] (3) [below of=2]{$q_3$};
    \node[state] (4) [below left of=3]{$q_4$};
    \node[state, accepting] (5) [left of=4]{$q_5$};
    \node[state] (6) [above left of=5]{$q_6$};
    \node[state] (7) [above of=6]{$q_7$};

    \path[->]
    (0) edge node [above]{$a_0$} (1)
    (0) edge node [right]{$a_1$} (5)
    (1) edge node [above right ]{$a_0$} (2)
    (1) edge node [right]{$a_1$} (6)
    (2) edge node [right]{$a_0$} (3)
    (2) edge node [pos=0.1, above]{$a_1$} (7)
    (3) edge node [below right]{$a_0$} (4) 
    (3) edge node [pos=0.45, left]{$a_1$} (0)
    (4) edge node [below]{$a_0$} (5)
    (4) edge node [left]{$a_1$} (1)
    (5) edge node [below left]{$a_0$} (6)
    (5) edge node [left]{$a_1$} (2)
    (6) edge node [left]{$a_0$} (7)
    (6) edge node [pos=0.1, above]{$a_1$} (3)
    (7) edge node [above left]{$a_0$} (0)
    (7) edge node [pos=0.6, above]{$a_1$} (4);
    \end{tikzpicture}}
\resizebox{0.6\textwidth}{!}{
    \begin{tikzpicture}[node distance=1.75cm, initial text = ] 

      \node[state,initial] (0000) {$\bit{0}, \bit{000}$};
    
      \node[state, right=of 0000]  (0001) {$\bit{0}, \bit{001}$};
      \node[state, right= of 0001] (0010) {$\bit{0}, \bit{010}$};
      \node[state, right= of 0010] (0011) {$\bit{0}, \bit{011}$};
      \node[state, below= 1cm of 0000]  (0100) {$\bit{0}, \bit{100}$};
      \node[state, right=of 0100]  (0101) {$\bit{0}, \bit{101}$};
      \node[state, right=of 0101]  (0110) {$\bit{0}, \bit{110}$};
      \node[state, right=of 0110]  (0111) {$\bit{0}, \bit{111}$};

      \node[state, below = 5cm of 0000] (1000) {$\bit{1}, \bit{000}$};
    
      \node[state, right=of 1000]  (1001) {$\bit{1}, \bit{001}$};
      \node[state, right= of 1001] (1010) {$\bit{1}, \bit{010}$};
      \node[state, right= of 1010] (1011) {$\bit{1}, \bit{011}$};
      \node[state, accepting, below= 1cm of 1000]  (1100) {$\bit{1}, \bit{100}$};
      \node[state, accepting, right=of 1100]  (1101) {$\bit{1}, \bit{101}$};
      \node[state, accepting, right=of 1101]  (1110) {$\bit{1}, \bit{110}$};
      \node[state, accepting, right=of 1110]  (1111) {$\bit{1}, \bit{111}$};
    
      \node[below left =0.5cm and 0.5cm of 0000]  (lineleft) {};
      \node[above right=0.5cm and 0.5cm of 1111] (lineright) {};
        
      \path[->] 
      (0000) edge node [above]{$b_0$} (0001)
      (0010) edge node [above]{$b_0$} (0011)
      (0100) edge node [above]{$b_0$} (0101)
      (0110) edge node [above]{$b_0$} (0111)
      (0000) edge [bend left]  node [above] {$b_1$} (0010)
      (0001) edge [bend left]  node [above] {$b_1$} (0011)
      (0100) edge [bend right]  node [below] {$b_1$} (0110)
      (0101) edge [bend right]  node [below] {$b_1$} (0111)
      (0000) edge node [left]{$b_2$} (0100)
      (0001) edge node [left]{$b_2$} (0101)
      (0010) edge node [left]{$b_2$} (0110)
      (0011) edge node [left]{$b_2$} (0111)
      
      (1001) edge [bend left]  node [above] {$a_1$} (1010)
      
      (1010) edge node [pos=0.1, above left] {$a_2$} (1100)
      
      (1011) edge node [pos=0.1, above left]{$a_2$} (1100)
      (1011) edge [bend right=30] node[above] {$a_1$} (1000)

      (1101) edge [bend right] node[below] {$a_1$} (1110)
      
      (1110) edge node[pos=0.1, above right] {$a_2$} (1000)
      
      (1111) edge node[pos=0.1, above right] {$a_2$} (1000)
      (1111) edge[bend left=30] node[below] {$a_1$} (1100)
 
      (0000) edge [bend right] node [left] {$r$} (1000)
      (0001) edge [bend right] node [pos=0.6, left] {$r$} (1001)
      (0010) edge [bend right] node [pos=0.3, above left] {$r$} (1010)
      (0011) edge [bend left] node [right] {$r$} (1011)
      (0100) edge [bend right] node [left] {$r$} (1100)
      (0101) edge [bend right] node [pos=0.6, left] {$r$} (1101)
      (0110) edge [bend left] node [pos=0.4, right] {$r$} (1110)
      (0111) edge [bend left] node [right] {$r$} (1111)
       ;
    \end{tikzpicture}}
    \end{center}
    \caption{The DFA $\C_5$ on the left, and the DFA $\mathcal{B}_3'$ (without self-loops) on the
    right.\label{fig:dfas-new}} 
\end{figure}

\paragraph{Cycle DFAs:}
This family has been used to benchmark our approach to DFA equivalence checking
using the GPU-accelerated model checker \GPUexplore. It is inspired by the Fibonacci DFAs,
but it differs in two aspects. First of all, the DFAs may involve multiple letters, to make equivalence checking more interesting. Second of all, the correspondence to Fibonacci words over the
binary alphabet has been removed, as it cannot be encoded in the input language of
\GPUexplore, \SLCO.

We define the sequence of Fibonacci numbers as follows: $\fib(0) = \fib(1) = 1$, and for all $n \in \Nat$ with $n>1$,
$\fib(n) = \fib(n-1) + \fib(n-2)$.
Now, for every $n\in \Nat$ with $n > 1$, we define the automaton $\C_n = (Q, \Sigma, \delta, q_0, F)$ as follows:
\begin{itemize}
  \item the set of states is $Q =\{q_i \mid i \in [0, \fib(n)-1 ] \}$; 
\item the set of actions $\Sigma$ consists of actions $a_0, \ldots, a_m$, with $m = \max(\lfloor \log_{10}(\fib(n)) \rceil, 1)$, i.e., the rounded logarithm of $\fib(n)$ or $1$, if that number is $0$.
\item the transition function is defined as follows, for $j \in [0, m]$:
\[
\delta(q_i,a_j) = q_{i+(j \cdot 100)+1\mod \fib(n)}
\]
  
  \item the set of final states is $F = \{q_{\fib(n-1)}\}$.
\end{itemize}

\noindent
The Cycle DFA $\C_5$ is given in Figure~\ref{fig:dfas-new}.

\paragraph{Extended bit-splitters:}
As \GPUexplore requires that input automata have an initial state, the bit-splitters can only be used
by our approach to DFA equivalence checking if they are extended with an initial state.
To achieve this, we extend the states in the bit-splitters to tuples $(\bit{c}, \sigma)$, with the first element being a
bit $\bit{c}$, and the second being the bit-sequences $\sigma$ of the original bit-splitters.
This effectively doubles the number of states of an extended bit-splitter $\mathcal{B}_n'$ compared
to $\mathcal{B}_n$.

We define 
$\mathcal{B}_1'= (Q_1',
\Sigma_1', \delta_1', (\bit0, \bit0), F_1')$, where $Q_1' = \{(\bit0, \bit0), (\bit0, \bit1), (\bit1, \bit0), (\bit1, \bit1)\}$ and $\Sigma_1' =
\{ b_0, r \}$. The letter $b_0$ occurs on transitions in which the second state bit is flipped from $\bit0$ to $\bit1$ and the first bit is $\bit0$.
In all other cases, $b_0$ occurs on a self-loop.
\[
    \delta_1'((\bit{c}, \bit{b}), b_0)= \begin{cases}
      (\bit{c}, \bit{\bar{b}}) & \text{If } \bit{c} = 0 \wedge \bit{b} = 0,\\
      (\bit{c}, \bit{b}) & \text{otherwise.}
    \end{cases}
\]
The letter $r$ occurs on transitions that set the first bit to $\bit1$:
\[
\delta_1'((\bit{c}, \bit{b}), r) = (\bit1, \bit{b})
\]
\noindent
Finally, $F_1'=\{(\bit1,\bit1)\}$.

Given the automaton $\mathcal{B}_n' = (Q_n', \Sigma_n',
\delta_n', F_n')$ for some $n \in \Nat$,
we define $\mathcal{B}_{n+1}' = (Q_{n+1}',
\Sigma_{n+1}', \delta_{n+1}', F_{n+1}')$, such that:
\begin{itemize}
  \item The set of states contains two copies of the $\sigma$ parts of the states in $Q_n'$,
  extended with one bit, combined with the two possible values of $\bit{c}$, i.e., $Q_{n+1}' = \{
  (\bit{c}, \bit0\sigma), (\bit{c}, \bit1\sigma) \mid \bit{c} \in [0,1] \wedge \sigma\in Q_n' \}$,
  \item Two fresh symbols $a_n, b_{n+1} \not\in \Sigma_n'$ are added to the alphabet: $\Sigma_{n+1}' = \Sigma_{n}'
  \cup \{ a_n, b_{n+1} \}$,
  \item The transition function $\delta_{n+1}'$ is defined such that for each
  $a_k, b_m \in \Sigma_n'$, and states $(\bit{c}, \bit{b}\sigma) \in Q_{n+1}'$, it maintains the
  behaviour of $\mathcal{B}_{n}'$, i.e., $\delta_{n+1}'((\bit{c}, \bit{b}\sigma), a_k) =
  (\bit{c}, \bit{b}\delta_n'((\bit{c}, \sigma), a_k))$ and $\delta_{n+1}'((\bit{c}, \bit{b}\sigma), b_m) =
  (\bit{c}, \bit{b}\delta_n'((\bit{c}, \sigma), b_m))$. For the fresh symbols $a_n, b_{n+1} \in \Sigma_{n+1}'
  \setminus \Sigma_n'$, $\delta_{n{+}1}'$ is extended as follows, with $\bit{\bar{b}}$ being
  the bit flipped version of $\bit{b}$, i.e., $\bit{b} = 0 \iff \bit{\bar{b}} = 1$:
  \[
    \delta_{n{+}1}'((\bit{c}, \bit{b}\sigma), a_n)= \begin{cases}
      (\bit{c}, \bit{\bar{b}} \bit0^n) & \text{If } \bit{c} = 1 \wedge \sigma[0] = 1,\\
      (\bit{c}, \bit{b}\sigma) & \text{otherwise.}
    \end{cases}
  \]
  \[
    \delta_{n{+}1}'((\bit{c}, \bit{b}\sigma), b_{n+1})= \begin{cases}
      (\bit{c}, \bit{\bar{b}} \sigma) & \text{If } \bit{c} = 0 \wedge \bit{b} = 0,\\
      (\bit{c}, \bit{b}\sigma) & \text{otherwise.}
    \end{cases}
  \]
\noindent
  Finally, $\delta_{n+1}$ is extended with $r$-transitions in the following way:
  \[
    \delta_{n{+}1}'((\bit{c}, \bit{b}\sigma), r)= (\bit{1}, \bit{b} \sigma)
  \]
  
  \item the set of accepting states is $F_{n+1}' = \{ (\bit1, \bit1 \sigma) \mid \sigma \in
  Q_n' \}$.
\end{itemize}

\noindent
In the extended bit-splitter family of DFAs, the $b_i$-transitions serve the purpose of manipulating 
the bits of the initial bit-sequence consisting of all $\bit0$'s, prior to entering the original bit-splitter
DFA, now represented by states in which $\bit{c} = \bit{1}$, and the $a_i$-transitions between
them. The $r$-transitions indicate moving from the `initialisation' part of the DFA to the
bit-splitter part of the DFA.
The extended bit-splitter $\mathcal{B}_3'$ is given in Figure~\ref{fig:dfas-new}, where, for
readability, the self-loops are not shown.

\begin{figure}
  \begin{center}
    \resizebox{0.49\textwidth}{!}{
  \begin{tikzpicture}[node distance= 1.5cm, initial text = ]
      \node[state,initial] (000) {$\bit{000}$};
    
      \node[state, right=of 000]  (001) {$\bit{001}$};
      \node[state, right= of 001] (010) {$\bit{010}$};
      \node[state, right= of 010] (011) {$\bit{011}$};
      \node[state, accepting, below= 1cm of 000]  (100) {$\bit{100}$};
      \node[state, accepting, right=of 100]  (101) {$\bit{101}$};
      \node[state, accepting, right=of 101]  (110) {$\bit{110}$};
      \node[state, accepting, right=of 110]  (111) {$\bit{111}$};
      
      \path[->]
      (000) edge [loop above] node{$0$} (000)
      (000) edge [bend left] node [above] {$1$} (001)
      (001) edge  node [above] {$0$} (010)
      (001) edge [bend left] node [above] {$1$} (011)
      (010) edge [pos=0.7, bend right=12]node [below] {$0$} (100)
      (010) edge[bend right=7]  node [above] {$1$} (101)
      (011) edge [bend left] node [below] {$0$} (110)
      (011) edge [bend left] node [right] {$1$} (111)
      (100) edge [bend left] node [left] {$0$} (000)
      (100) edge [bend left] node [above] {$1$} (001)
      (101) edge [bend right=7] node [below] {$0$} (010)
      (101) edge [pos=0.7, bend right=12] node [above] {$1$} (011)
      (110) edge [bend left] node [below] {$0$} (100)
      (110) edge  node [below] {$1$} (101)
      (111) edge [bend left] node [below] {$0$} (110)
      (111) edge [loop below] node{$1$} (111);
    \end{tikzpicture}}
    \resizebox{0.49\textwidth}{!}{
  \begin{tikzpicture}[node distance= 1.5cm, initial text = ]
      \node[state,initial] (000) {$\bit{000}$};
    
      \node[state, right=of 000]  (001) {$\bit{001}$};
      \node[state, right= of 001] (010) {$\bit{010}$};
      \node[state, right= of 010] (011) {$\bit{011}$};
      \node[state, accepting, below= 1cm of 000]  (100) {$\bit{100}$};
      \node[state, accepting, right=of 100]  (101) {$\bit{101}$};
      \node[state, accepting, right=of 101]  (110) {$\bit{110}$};
      \node[state, accepting, right=of 110]  (111) {$\bit{111}$};
      
      \path[->]
      (000) edge [loop above] node{$0$} (000)
      (000) edge [bend left] node [above] {$1$} (001)
      (001) edge  node [above] {$0$} (010)
      (001) edge [bend left] node [above] {$1$} (011)
      (010) edge [pos=0.9, bend right=12]node [below] {$0$} (100)
      (010) edge node [below] {$1$} (101)
      (011) edge [bend left] node [below] {$0$} (110)
      (011) edge [bend left] node [right] {$1$} (111)
      (100) edge [bend left] node [left] {$0$} (000)
      (100) edge [pos=0.7, bend left] node [above] {$1$} (001)
      (101) edge [pos=0.2, bend right=7] node [below] {$0$} (000)
      (101) edge [pos=0.3] node [right] {$1$} (001)
      (110) edge [bend left] node [below] {$0$} (100)
      (110) edge  node [below] {$1$} (101)
      (111) edge [bend left] node [below] {$0$} (110)
      (111) edge [loop below] node{$1$} (111);
    \end{tikzpicture}}
    \end{center}
    \caption{The perfect (left) and forgetful (right) $\memory_3$ DFAs.\label{fig:memory-perfect}} 
\end{figure}
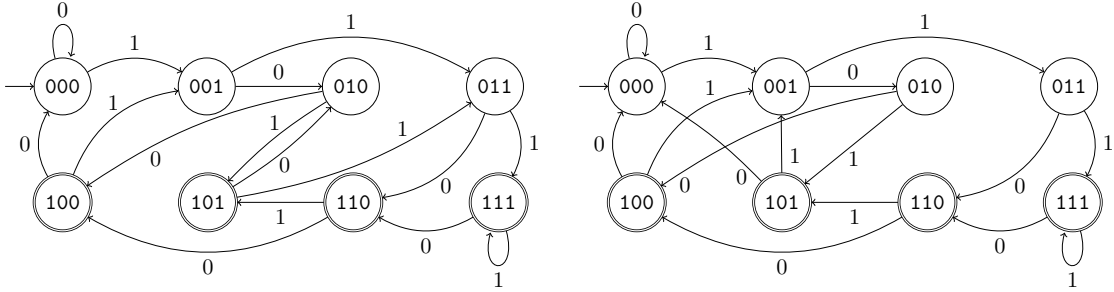

\paragraph{Memory}
The $\memory_n$ DFA family is used to perform inclusion checking. For each depth $n \in \mathbb{N}$ two members of this family are defined, referred to as \textit{perfect} and \textit{forgetful}. For both of these, each state of the DFA for depth $n$ encodes information about the last $n$ symbols the DFA has read. The alphabet we use only contains two elements, so each state represents a unique bitstring of length $n$. For each symbol in the alphabet an outgoing transition with that symbol as the action is defined to the state representing the bitstring of the source state with the respective bit appended. The most significant bit is discarded. In the forgetful memory DFA all bits are reset after the most significant bits read $10$.
The \textit{perfect} $\memory$ DFA for depth $n=2$, i.e., $\memory_2$, can be seen in Figure~\ref{fig:mem2}, as it
is used to explain the encoding of DFA equivalence checking in \SLCO models.
The \textit{perfect} and \textit{forgetful} $\memory$ DFAs for depth $n=3$ are displayed in Figure~\ref{fig:memory-perfect}.

\subsection{Results for DFA minimisation}
\label{sec:results}
The DFA minimisation algorithms were implemented in CUDA \CC and compiled using the CUDA toolkit 12.2, with
the implementation of \texttt{sortPR} using the Thrust library for sorting and
computing the adjacent differences and inclusive scans~\cite{thrust}.
Experiments were conducted on a device running Linux Mint 20, equipped with an
NVIDIA TITAN RTX GPU with 24 GB of memory and 4,608 cores. Such a GPU can manage
trillions of lightweight threads. Thanks to fast context switching between threads, a GPU can
 typically handle a few hundred thousand threads as if they execute in parallel. 

The reported times are the average of five separate runs.
Each time includes the time it takes to copy the
problem instance to the GPU memory.
Benchmarks that did
not finish within five minutes were aborted, in which case we registered a
timeout `t/o'. Benchmarks for which there was not enough memory are indicated by
`OoM'.

\begin{table}[tb]
  \centering
  \begin{tabular}{lcrrrr}
    \toprule
    Name & $N$ & Iterations & Time (ms) & Memory(Mb) & \#threads \\
    \midrule
    $\Fib_4$ & 8 & 3 & 0.3 & 0 & 589,824\\
    $\Fib_5$ & 13 & 4 & 0.7 & 0 & 6,230,016\\
    $\Fib_6$ & 21 & 5 & 7.8 & 0 & 88,510,464\\
    $\Fib_7$ & 34 & 5 & 159.9 & 0 & 1,620,545,536\\
    $\Fib_8$ & 55 & 6 & 3,034.9 & 10 & 27,955,840,000\\
    $\Fib_9$  & 89 & 7 & 66,846.7 & 60 & 498,865,340,416\\
    $\Fib_{10}$ & 144 & t/o & t/o & 412 & 8,943,640,510,464\\
    \bottomrule
  \end{tabular}
  \caption{Results of running the algorithm $\texttt{trans}$ on the Fibonacci automata.\label{table:trans}}
  \end{table}

\paragraph{Transitive approach:}
The results of running Algorithm~\ref{alg:fulltrans} on the Fibonacci automata
are given in Table~\ref{table:trans}. As expected, the number of threads used to
compute the transitive closure in parallel grows very quickly. Although the
number of iterations of the algorithm is indeed logarithmic, the available
parallelism is not sufficient to lead to logarithmic run times. We only use
this set of small Fibonacci automata for this algorithm. It already suggests
that for a relatively small amount of states (~$\sim 100$), obtaining the required resources is
already infeasible. The other benchmarks are almost completely out of range of
the algorithm.

  \begin{table}[t]
    \centering
    \resizebox{\textwidth}{!}{
  \begin{tabular}{l rrr rrr rrr}
    \toprule
    & \multicolumn{3}{c}{\textbf{Benchmark metrics}} &
      \multicolumn{3}{c}{\textbf{Times (ms)}} &
      \multicolumn{3}{c}{\textbf{Iterations}}\\
    \cmidrule(lr){2-4} \cmidrule(lr){5-7} \cmidrule(lr){8-10}
    Name & \multicolumn{1}{c}{$N$} & \multicolumn{1}{c}{$|\Sigma|$} &
           \multicolumn{1}{c}{Size output} & \multicolumn{1}{c}{$\mathtt{naivePR}$} &
           \multicolumn{1}{c}{$\mathtt{sortPR}$} & \multicolumn{1}{c}{$\mathtt{transPR}$} &
           \multicolumn{1}{c}{$\mathtt{naivePR}$} & \multicolumn{1}{c}{$\mathtt{sortPR}$} &
           \multicolumn{1}{c}{$\mathtt{transPR}$}\\
    \midrule
  $\Fib_{18}$        & 6,765      & 1  & 6,765                          & 118.6                       & 979.0     & 1.0     & 6,764    & 6,764                    & 5                       \\
  $\Fib_{19}$        & 10,946     & 1  & 10,946                         & 191.7                       & 2,085.9   & 1.1     & 10,945   & 10,945                   & 12                      \\
  $\Fib_{20}$        & 17,711     & 1  & 17,711                         & 308.8                       & 3,909.2   & 1.7     & 17,710   & 17,710                   & 14                      \\
  $\Fib_{21}$        & 28,657     & 1  & 28,657                         & 494.2                       & 6,374.2   & 2.4     & 28,656   & 28,656                   & 25                      \\
  $\Fib_{22}$        & 46,368     & 1  & 46,368                         & 778.7                       & 11,712.1  & 4.1     & 46,367   & 46,367                   & 61                      \\
  $\Fib_{23}$        & 75,025     & 1  & 75,025                         & 1,241.3                     & 21,366.6  & 8.0     & 75,024   & 75,024                   & 101                     \\
  $\Fib_{24}$        & 121,393    & 1  & 121,393                        & 2,006.7                     & 34,793.1  & 12.5    & 121,392  & 121,392                  & 104                     \\
  $\Fib_{25}$        & 196,418    & 1  & 196,418                        & 3,251.3                     & 64,411.7  & 18.3    & 196,417  & 196,417                  & 138                     \\
  $\Fib_{26}$        & 317,811    & 1  & 317,811                        & 5,277.8                     & 178,367.4 & 49.8    & 317,810  & 317,810                  & 102                     \\
  $\Fib_{27}$        & 514,229    & 1  & 514,229                        & 8,607.7                     & t/o       & 96.1    & 514,228  & \multicolumn{1}{r}{t/o} & 268                     \\
  $\Fib_{28}$        & 832,040    & 1  & 832,040                        & 22,723.0                    & t/o       & 178.4   & 832,039  & \multicolumn{1}{r}{t/o} & 299                     \\
  $\Fib_{29}$        & 1,346,269  & 1  & 1,346,269                      & 59,510.8                    & t/o       & 726.9   & 1,346,268 & \multicolumn{1}{r}{t/o} & 755                     \\
  $\Fib_{30}$        & 2,178,309  & 1  & 2,178,309                      & 141,601.0                   & t/o       & 1,109.3 & 2,178,308 & \multicolumn{1}{r}{t/o} & 914                     \\
  \midrule
  $\mathcal{B}_{15}$ & 32,768     & 14 & 32,768                         & 0.8                         & 25.8      & 1.7     & 14      & 14                      & 2                       \\
  $\mathcal{B}_{16}$ & 65,536     & 15 & 65,536                         & 1.4                         & 29.7      & 3.7     & 15      & 15                      & 2                       \\
  $\mathcal{B}_{17}$ & 131,072    & 16 & 131,072                        & 2.6                         & 54.3      & 9.4     & 16      & 16                      & 2                       \\
  $\mathcal{B}_{18}$ & 262,144    & 17 & 262,144                        & 5.0                         & 107.2     & 25.6    & 17      & 17                      & 2                       \\
  $\mathcal{B}_{19}$ & 524,288    & 18 & 524,288                        & 9.6                         & 235.7     & 60.9    & 18      & 18                      & 2                       \\
  $\mathcal{B}_{20}$ & 1,048,576  & 19 & 1,048,576                      & 19.3                        & 520.2     & 139.8   & 19      & 19                      & 2                       \\
  $\mathcal{B}_{21}$ & 2,097,152  & 20 & 2,097,152                      & 39.8                        & 1,148.6   & 312.2   & 20      & 20                      & 2                       \\
  $\mathcal{B}_{22}$ & 4,194,304  & 21 & 4,194,304                      & 82.6                        & 2,538.5   & 728.7   & 21      & 21                      & 2                       \\
  $\mathcal{B}_{23}$ & 8,388,608  & 22 & 8,388,608                      & 170.3                       & 5,612.7   & 1,612.1 & 22      & 22                      & 2                       \\
  $\mathcal{B}_{24}$ & 16,777,216 & 23 & 16,777,216                     & 352.6                       & 12,351.8  & OoM     & 23      & 23                      & OoM                     \\
  $\mathcal{B}_{25}$ & 33,554,432 & 24 & 33,554,432                     & 737.4                       & 27,092.2  & OoM     & 24      & 24                      & OoM \\
  $\mathcal{B}_{26}$ & 67,108,864 & 25 & 67,108,864                     & 1,541.5                     & 59,203.8  & OoM     & 25      & 25                      & OoM\\
    \bottomrule
  \end{tabular}}
  \caption{Results of running the partition refinement algorithms on the $\Fib$ and $\mathcal{B}$ benchmarks.\label{table:concur}}
  \end{table}

\paragraph{Partition refinement algorithms:} The results of running the parallel
partition refinement algorithms,
$\texttt{naivePR}$~(Algorithm~\ref{alg:partref}),
$\texttt{sortPR}$~(Algorithm~\ref{alg:sort}), and
$\texttt{transPR}$~(Algorithm~\ref{alg:partref-trans}) are given in
Table~\ref{table:concur} and Table~\ref{table:vlts} for the different
benchmarks.

First, we observe in Table~\ref{table:concur} that on the Fibonacci automata the
$\texttt{naivePR}$ performs better than $\texttt{sortPR}$. This can be explained
by the fact that the number of iterations is $n$ for both algorithms, while each
iteration in $\texttt{sortPR}$ is slower than in $\texttt{naivePR}$. Another
interesting observation here is that for all benchmarks $\Fib_{18}, \dots,
\Fib_{28}$ the run time of the algorithm $\texttt{naivePR}$ scales linearly with
the number of states $n$. Since the number of parallel iterations is $n{-}2$ for
all these benchmarks, each parallel iteration processing up to $\sim 500k$
states took a similar amount of time. In other words, the GPU was able to run
around $500k$ threads as if they ran in parallel. This confirms the statement
about fast context switching at the beginning of Section~\ref{sec:results}.

Finally, for the Fibonacci automata, we see that $\texttt{transPR}$ performs
significantly better on this benchmark. This can be explained by the fact that the
partial transitive closure reduces the number of iterations of the algorithm
significantly. 

The results on the bit-splitter automata in Table~\ref{table:concur} show that
the improvement of $\texttt{transPR}$ does not work on all automata. The high
number of alphabet letters together with the structure of the automata make the
transitive closure less effective, making $\texttt{naivePR}$ much faster.

For the VLTS benchmark set we see the power of $\texttt{sortPR}$ in Table~\ref{table:vlts}. In some
benchmarks, like `vasy\_69\_520' the algorithm performs significantly better. In
these examples, it helps that in each iteration of $\texttt{sortPR}$, a block
can be split into many subblocks, which is not the case in the other algorithms.

Since the VLTS benchmarks originate from communication
protocols and concurrent systems, the success of \texttt{sortPR} suggests that
for DFAs that represent `real' systems, this algorithm is a solid choice for
efficient DFA minimisation. However, the experiments with the Fibonacci and
bit-splitter families of DFAs demonstrate room for improvement.


\begin{table}[tb]
  \centering
  \resizebox{\textwidth}{!}{
  \begin{tabular}{l rrr rrr rrr}
    \toprule
    & \multicolumn{3}{c}{\textbf{Benchmark metrics}} &
      \multicolumn{3}{c}{\textbf{Times (ms)}} &
      \multicolumn{3}{c}{\textbf{Iterations}}\\
    \cmidrule(lr){2-4} \cmidrule(lr){5-7} \cmidrule(lr){8-10}
    Name &
    \multicolumn{1}{c}{$N$} & \multicolumn{1}{c}{$|\Sigma|$} & \multicolumn{1}{c}{Size output} &
    \multicolumn{1}{c}{$\mathtt{naivePR}$} & \multicolumn{1}{c}{$\mathtt{sortPR}$} & \multicolumn{1}{c}{$\mathtt{transPR}$} &
    \multicolumn{1}{c}{$\mathtt{naivePR}$} & \multicolumn{1}{c}{$\mathtt{sortPR}$} & \multicolumn{1}{c}{$\mathtt{transPR}$}\\
    \midrule
    cwi\_1\_2         & 4,448     & 26     & 2,416   & 5.4       & 66.7     & 25.1      & 308    & 38    & 621   \\
    cwi\_2416\_17605  & 503       & 15     & 58      & 0.8       & 38.2     & 0.4       & 40     & 40    & 8     \\
    cwi\_3\_14        & 63        & 2      & 63      & 1.2       & 9.1      & 0.4       & 61     & 61    & 8     \\
    vasy\_0\_1        & 92        & 2      & 10      & 0.2       & 3.9      & 0.4       & 6      & 5     & 5     \\
    vasy\_1\_4        & 6,087     & 6      & 29      & 0.4       & 8.5      & 0.9       & 15     & 7     & 20    \\
    vasy\_10\_56      & 10,850    & 12     & 2113    & 8.7       & 40.2     & 30.9      & 519    & 33    & 791   \\
    vasy\_1112\_5290  & 1,112,491 & 23     & 266     & 135.4     & 386.8    & 2,049.2   & 246    & 4     & 231   \\
    vasy\_157\_297    & 157,605   & 235    & 4,290   & 455.1     & 1,736.3  & 11,312.0  & 1,049   & 27    & 1,306  \\
    vasy\_164\_1619   & 109,911   & 37     & 1,025   & 69.9      & 50.5     & 823.4     & 770    & 4     & 766   \\
    vasy\_166\_651    & 393,147   & 211    & 392,175 & 159,265.6 & 1,070.6  & t/o       & 175,764 & 19    & t/o   \\
    vasy\_18\_73      & 419,664   & 17     & 31,952  & 1,586.1   & 305.2    & 34,055.2  & 13,343  & 27    & 18,444 \\
    vasy\_25\_25      & 25,218    & 25,216 & 25,218  & 262,878.6 & 3,502.7  & t/o       & 25,217  & 2     & t/o   \\
    vasy\_386\_1171   & 355,790   & 73     & 114     & 36.9      & 489.4    & 766.0     & 58     & 8     & 113   \\
    vasy\_40\_60      & 40,007    & 3      & 40,007  & 331.6     & 8,391.5  & 845.2     & 20,004  & 20,002 & 20,004 \\
    vasy\_5\_9        & 5,088     & 31     & 138     & 2.2       & 14.3     & 7.0       & 113    & 5     & 124   \\
    vasy\_574\_13561  & 574,058   & 141    & 3,578   & 2,332.2   & 976.5    & 64,312.6  & 2,351   & 5     & 2,634  \\
    vasy\_6120\_11031 & 3,190,785 & 125    & 5,216   & 13,186.6  & 21,886.0 & t/o       & 2,373   & 21    & t/o   \\
    vasy\_65\_2621    & 65,538    & 72     & 65,537  & 2,591.8   & 38.3     & 47,568.0  & 36,575  & 4     & 38,999 \\
    vasy\_66\_1302    & 209,791   & 81     & 208,419 & 42,864.9  & 96.0     & t/o       & 179,861 & 8     & t/o   \\
    vasy\_69\_520     & 74,958    & 135    & 74,958  & 7,223.0   & 124.2    & 181,611.4 & 49,723  & 12    & 74,667 \\
    vasy\_720\_390    & 87,741    & 49     & 3,279   & 176.0     & 57.1     & 2,961.7   & 2,936   & 5     & 2,950  \\
    vasy\_8\_24       & 20,306    & 11     & 560     & 5.9       & 26.8     & 22.1      & 282    & 17    & 348   \\
    vasy\_8\_38       & 8,922     & 81     & 220     & 5.7       & 44.1     & 31.5      & 174    & 5     & 215   \\
    vasy\_83\_325     & 393,147   & 211    & 392,175 & 162,495.0 & 1,074.4  & t/o     & 173,218 & 19    & t/o\\
    \bottomrule
    \end{tabular}}
  \caption{Results of running the partition refinement algorithms on the VLTS benchmark set.\label{table:vlts}}
  \end{table}

\subsection{Results for DFA equivalence and inclusion checking.}
As explained in Section~\ref{sec:dfa-eq}, high-level descriptions of the two DFAs involved in
equivalence or inclusion checking are written in the \SLCO model checking language before the state space of their synchronous product is explored using the \GPUexplore model checking tool. Depending on the model checking task, \GPUexplore reports multiple timing metrics related to the running time of the state space exploration task. Each time such a task is initiated, a CUDA context needs to be initialised, including moving the problem instance to the GPU.
This takes up to a second. We refer to the time this takes as the \textit{initialisation time}. The time spent on actually exploring the state space is called the
 \textit{exploration time}, which is measured separately. In theory it should be possible to eliminate 
 most of the initialisation time when running multiple exploration tasks by using the same CUDA 
 context, but we have not implemented this functionality.

We have tested the benchmarks on two different systems, to assess how much the hardware configuration affects the performance of the algorithm implementations.
One system has an Intel i5\texttrademark\ 6600k processor with 32GB of DDR4 RAM. The system is equipped with two NVIDIA Titan RTX graphics cards, each containing 4,608 CUDA cores and 24GB of GDDR6 VRAM running with a memory clock of 1,750 MHz, at a $384$-bit bus width. We only use one card at a time for these benchmarks.

The other system has an AMD Ryzen\texttrademark\ 7 5800X, installed on an MSI MPG x570 Gaming Pro Carbon Wi-Fi motherboard, with 32GB of DDR4 RAM. This system is equipped with an NVIDIA RTX 3090 GPU, with 10,496 CUDA cores and 24GB of GDDR6X VRAM running with a memory clock of 1,219 MHz, at a $384$-bit bus width. 

\begin{table}[t]
  \centering
  \resizebox{0.735\textwidth}{!}{
  \begin{tabular}{l r rrr rrr}
    \toprule
    & \multicolumn{1}{c}{\textbf{Benchmark metrics}} &
      \multicolumn{3}{c}{\textbf{RTX 3090 Times (ms)}} &
      \multicolumn{3}{c}{\textbf{Titan RTX Times (ms)}}\\
    \cmidrule(lr){2-2} \cmidrule(lr){3-5} \cmidrule(lr){6-8}
    Name &
    \multicolumn{1}{c}{State count} &
    \multicolumn{1}{c}{Total} & \multicolumn{1}{c}{Initial} & \multicolumn{1}{c}{Exploration} &
    \multicolumn{1}{c}{Total} & \multicolumn{1}{c}{Initial} & \multicolumn{1}{c}{Exploration}\\
    \midrule
$\B_{5}'$     & 64            & 274     & 274 & 0       & 592       & 591 & 0         \\
$\B_{6}'$     & 128           & 244     & 244 & 0       & 478       & 477 & 0         \\
$\B_{7}'$     & 256           & 243     & 243 & 0       & 479       & 478 & 0         \\
$\B_{8}'$     & 512           & 245     & 244 & 1       & 483       & 482 & 1         \\
$\B_{9}'$     & 1,024         & 243     & 243 & 1       & 467       & 466 & 1         \\
$\B_{10}'$    & 2,048         & 256     & 248 & 8       & 485       & 475 & 10        \\
$\B_{11}'$    & 4,096         & 269     & 246 & 23      & 494       & 464 & 30        \\
$\B_{12}'$    & 8,192         & 284     & 244 & 40      & 533       & 472 & 60        \\
$\B_{13}'$    & 16,384        & 298     & 242 & 56      & 530       & 460 & 70        \\
$\B_{14}'$    & 32,768        & 278     & 240 & 38      & 502       & 456 & 47        \\
$\B_{15}'$    & 65,536        & 279     & 238 & 41      & 517       & 465 & 52        \\
$\B_{16}'$    & 131,072       & 291     & 245 & 46      & 518       & 458 & 60        \\
$\B_{17}'$    & 262,144       & 290     & 237 & 53      & 529       & 460 & 68        \\
$\B_{18}'$    & 524,288       & 301     & 240 & 61      & 548       & 469 & 80        \\
$\B_{19}'$    & 1,048,576     & 310     & 237 & 73      & 582       & 470 & 112       \\
$\B_{20}'$    & 2,097,152     & 367     & 237 & 130     & 767       & 465 & 302       \\
$\B_{21}'$    & 4,194,304     & 578     & 245 & 333     & 1,517     & 453 & 1,063     \\
$\B_{22}'$    & 8,388,608     & 955     & 239 & 717     & 2,951     & 460 & 2,491     \\
$\B_{23}'$    & 16,777,216    & 1,997   & 240 & 1,757   & 6,347     & 460 & 5,887     \\
$\B_{24}'$    & 33,554,432    & 4,308   & 243 & 4,065   & 13,403    & 456 & 12,947    \\
$\B_{25}'$    & 67,108,864    & 9,467   & 240 & 9,227   & 28,490    & 451 & 28,039    \\
$\B_{26}'$    & 134,217,728   & 21,326  & 243 & 21,083  & 63,432    & 480 & 62,951    \\
$\B_{27}'$    & 268,435,456   & 48,648  & 242 & 48,407  & 143,937   & 459 & 143,478   \\
$\B_{28}'$    & 536,870,912   & 109,184 & 247 & 108,936 & 325,268   & 466 & 324,802   \\
$\B_{29}'$    & 1,073,741,824 & 247,849 & 241 & 247,607 & 732,229   & 464 & 731,764   \\
\midrule
$\C_{30}$     & 1,346,269     & 521     & 282 & 239     & 844       & 572 & 271       \\
$\C_{31}$     & 2,178,309     & 508     & 158 & 350     & 809       & 438 & 371       \\
$\C_{32}$     & 3,524,578     & 717     & 161 & 556     & 989       & 436 & 553       \\
$\C_{33}$     & 5,702,887     & 1,019   & 162 & 857     & 1,324     & 441 & 883       \\
$\C_{34}$     & 9,227,465     & 1,587   & 157 & 1,430   & 1,836     & 442 & 1,394     \\
$\C_{35}$     & 14,930,352    & 2,370   & 157 & 2,212   & 2,755     & 430 & 2,324     \\
$\C_{36}$     & 24,157,817    & 3,815   & 159 & 3,655   & 3,749     & 441 & 3,307     \\
$\C_{37}$     & 39,088,169    & 5,463   & 165 & 5,297   & 6,002     & 436 & 5,564     \\
$\C_{38}$     & 63,245,986    & 8,845   & 167 & 8,676   & 9,739     & 429 & 9,308     \\
$\C_{39}$     & 102,334,155   & 14,665  & 160 & 14,502  & 14,530    & 441 & 14,085    \\
$\C_{40}$     & 165,580,141   & 24,170  & 160 & 24,005  & 24,866    & 437 & 24,424    \\
$\C_{41}$     & 267,914,296   & 40,477  & 157 & 40,313  & 41,525    & 445 & 41,071    \\
$\C_{42}$     & 433,494,437   & 61,922  & 160 & 61,752  & 64,497    & 435 & 64,049    \\
    \bottomrule
\end{tabular}}
  \caption{Results of running \GPUexplore on the $\mathcal{B}'_n$ and $\C_n$ benchmarks.\label{table:gpuexplore-equivalence}}
  \end{table}

The state count of each state space along with all results are displayed in Table~\ref{table:gpuexplore-equivalence}. For all models in this table, the equivalence checking is trivial, i.e. we compare a DFA with itself. This means that the size of the resulting product is equivalent to the size of the input DFA, which means that worst-case, the running time is linear. While there are some irregularities for lower state counts, the graphs for extended bit-splitter and Cycle DFAs support this proposition.

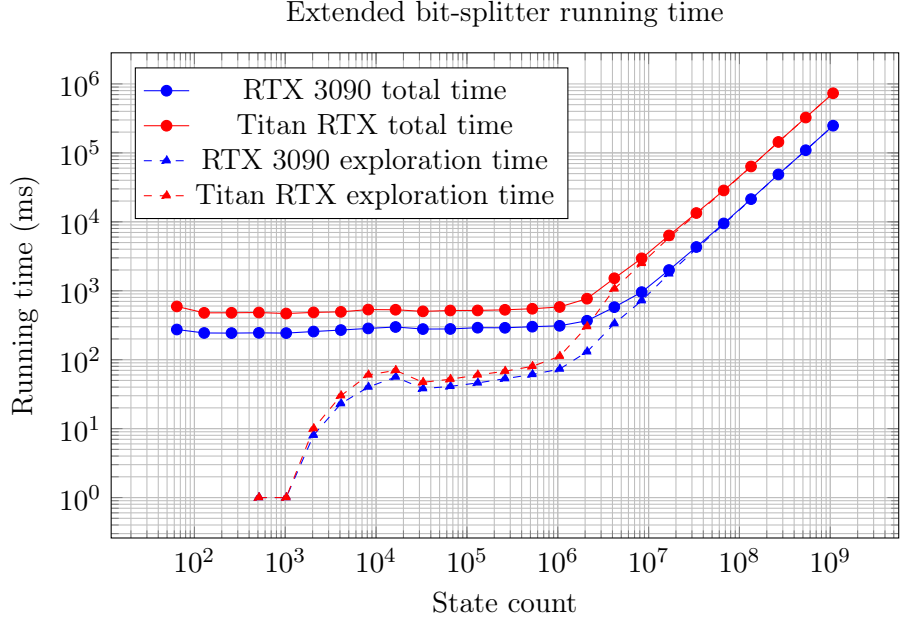
\begin{figure}[t!]
\begin{center}
\begin{tikzpicture}
\begin{axis}[
    xmode=log,
    ymode=log,
    width=12cm,
    height=8cm,
    xlabel={State count},
    ylabel={Running time (ms)},
    title={Extended bit-splitter running time},
    grid=both,
    legend pos=north west
]

\addplot[
    color=blue,
    mark=*,
]
table[
    col sep=comma,
    x expr=\thisrowno{1},
    y expr=\thisrowno{2}
] {bitsplitter.csv};
\addlegendentry{RTX 3090 total time}

\addplot[
    color=red,
    mark=*,
]
table[
    col sep=comma,
    x expr=\thisrowno{1},
    y expr=\thisrowno{5}
] {bitsplitter.csv};
\addlegendentry{Titan RTX total time}

\addplot[
    dashed,
    color=blue,
    mark=triangle*,
]
table[
    col sep=comma,
    x expr=\thisrowno{1},
    y expr=\thisrowno{4}
] {bitsplitter.csv};
\addlegendentry{RTX 3090 exploration time}

\addplot[
    dashed,
    color=red,
    mark=triangle*,
]
table[
    col sep=comma,
    x expr=\thisrowno{1},
    y expr=\thisrowno{7}
] {bitsplitter.csv};
\addlegendentry{Titan RTX exploration time}

\end{axis}
\end{tikzpicture}
\end{center}

\caption{Exploration time and total time for different configurations of $\mathcal{B}'_n$, ranging from $n=5$ up to and including $n=29$. The total time includes the initialisation time of the CUDA context.}
\label{bitsplitter-graph}

\end{figure}

The results for the extended bit-splitter experiments are presented in Figure~\ref{bitsplitter-graph}. In the lower range, the exploration time of equivalence checking of extended bit-splitter DFAs follows a peculiar trajectory that we have found difficult to explain. Up to $n=13$ the exploration time increases very sharply, but from $n=13$ to $n=14$ the exploration time actually decreases. This pattern occurs for both the Titan RTX and the RTX 3090 systems. We believe this is related to the data type used for describing the individual states in the synchronous product, i.e., the elements
stored in the global hash table. For $n$ up to $13$, a 32-bit integer suffices to represent a state, but starting with $n=14$, a 64-bit integer is required for this. While one might expect the performance to suffer from the size increase, we hypothesise that better chunk alignment can actually increase the performance of memory accesses when using larger datatypes, and since a 
size of $64$ bits is still well within the bus width of the GPUs, it is to be expected that the increased 
size does not decrease performance of any other operations.


\begin{figure}
\begin{center}
\begin{tikzpicture}
\begin{axis}[
    xmode=log,
    ymode=log,
    width=12cm,
    height=8cm,
    xlabel={State count},
    ylabel={Running time (ms)},
    title={Cycle DFA running time},
    grid=both,
    legend pos=north west
]

\addplot[
    color=blue,
    mark=*,
]
table[
    col sep=comma,
    x expr=\thisrowno{1},
    y expr=\thisrowno{2}
] {multifib.csv};
\addlegendentry{RTX 3090 total time}

\addplot[
    color=red,
    mark=*,
]
table[
    col sep=comma,
    x expr=\thisrowno{1},
    y expr=\thisrowno{5}
] {multifib.csv};
\addlegendentry{Titan RTX total time}

\end{axis}
\end{tikzpicture}
\end{center}

\caption{Total time for different configurations of $\C_n$, ranging from $n=30$ up to and including $n=42$. The running time is not shown separately because it did not differ significantly from the total runtime, and it made the graph too cluttered.}
\label{multifib-graph}

\end{figure}
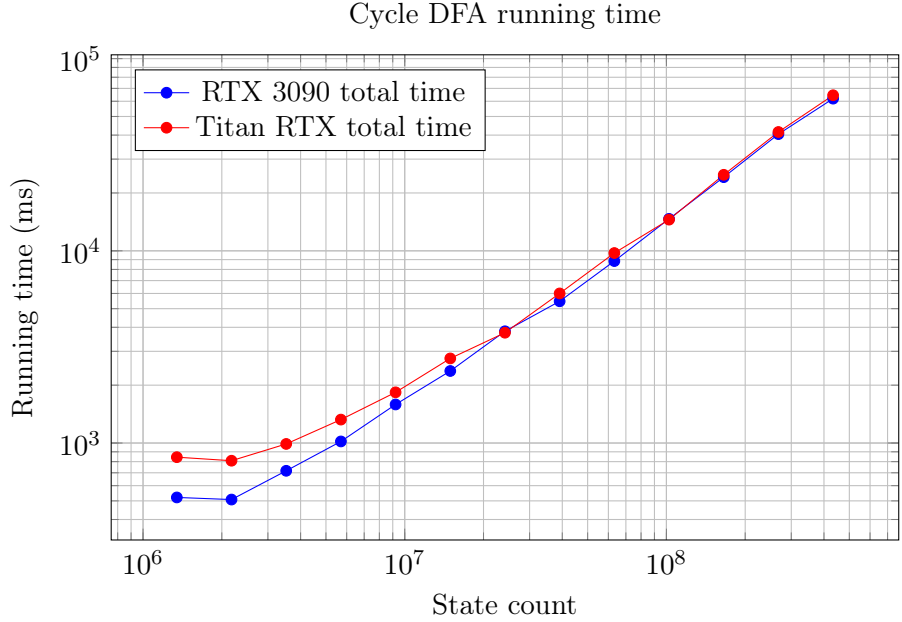

The results for the Cycle DFA experiments are shown in Figure~\ref{multifib-graph}. Where the RTX 3090 clearly outperformed the Titan RTX for the bit-splitter experiments, for the Cycle DFAs we can see that their times are a lot closer. For some models, the Titan RTX even beats the RTX 3090. While most hardware specifications are higher for the RTX 3090, the memory clock speed of the Titan RTX is higher than that of the RTX 3090. We suspect the performance of equivalence checking of Cycle DFAs 
may be bottlenecked by the memory throughput and frequency.

We used models for the perfect and forgetful $\memory_n$ DFAs to perform inclusion checking. Since the DFAs we compare are not equivalent, the resulting product can grow larger than the input DFAs, up to quadratic size. When exploring the state space of the $\memory$ model instances, the increase in states over the base DFAs was noticeable but not dramatic.

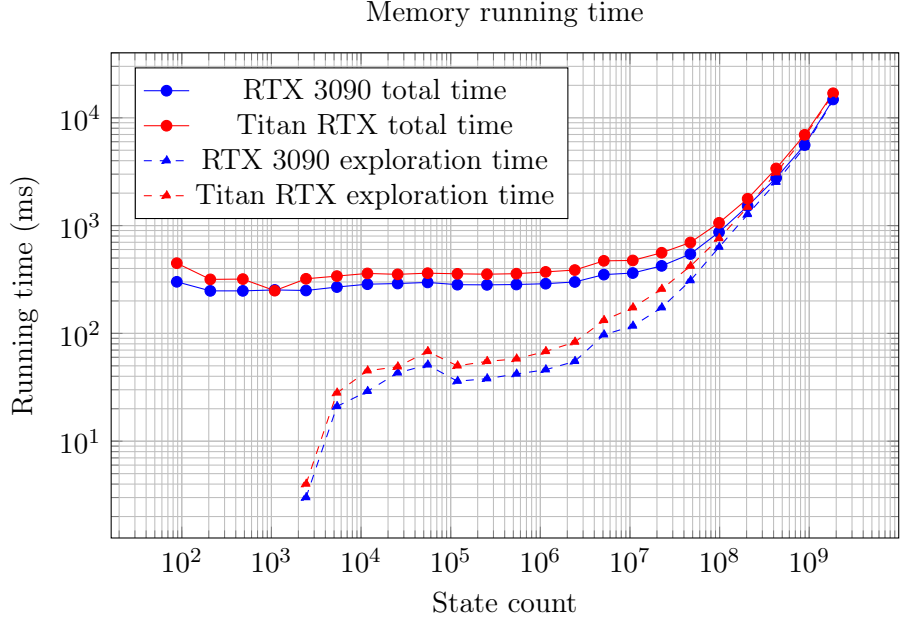
\begin{figure}
\begin{center}
\begin{tikzpicture}
\begin{axis}[
    xmode=log,
    ymode=log,
    width=12cm,
    height=8cm,
    xlabel={State count},
    ylabel={Running time (ms)},
    title={Memory running time},
    grid=both,
    legend pos=north west
]

\addplot[
    color=blue,
    mark=*,
]
table[
    col sep=comma,
    x expr=\thisrowno{1},
    y expr=\thisrowno{2}
] {memory.csv};
\addlegendentry{RTX 3090 total time}

\addplot[
    color=red,
    mark=*,
]
table[
    col sep=comma,
    x expr=\thisrowno{1},
    y expr=\thisrowno{5}
] {memory.csv};
\addlegendentry{Titan RTX total time}

\addplot[
    dashed,
    color=blue,
    mark=triangle*,
]
table[
    col sep=comma,
    x expr=\thisrowno{1},
    y expr=\thisrowno{4}
] {memory.csv};
\addlegendentry{RTX 3090 exploration time}

\addplot[
    dashed,
    color=red,
    mark=triangle*,
]
table[
    col sep=comma,
    x expr=\thisrowno{1},
    y expr=\thisrowno{7}
] {memory.csv};
\addlegendentry{Titan RTX exploration time}

\end{axis}
\end{tikzpicture}
\end{center}

\caption{Exploration time and total time for different configurations of $\memory_n$, ranging from $n=5$ up to and including $n=27$. The total time includes the initialisation time of the CUDA context.}
\label{memory-graph}

\end{figure}

The benchmark checks if the language of the forgetful $\memory$ model is included in the language of the perfect $\memory$ model. This inclusion is valid, as forgetting the state and resetting to the initial state only rejects a set of otherwise accepting inputs, and does not lead
to accepting inputs that would otherwise have been rejected. The results for this are shown in Table~\ref{table:inclusion-checking}, and a graph of the metrics is provided in Figure~\ref{memory-graph}.
The exploration times for the $\memory$ model instances are typically very low, so it is hard to draw conclusions about the overall running times from the graph. We were not able to perform inclusion checking for larger DFAs as our GPUs ran out of memory in those cases. 

\begin{table}
  \centering
  \resizebox{0.735\textwidth}{!}{
  \begin{tabular}{l r rrr rrr}
    \toprule
    & \multicolumn{1}{c}{\textbf{Benchmark metrics}} &
      \multicolumn{3}{c}{\textbf{RTX 3090 Times (ms)}} &
      \multicolumn{3}{c}{\textbf{Titan RTX Times (ms)}}\\
    \cmidrule(lr){2-2} \cmidrule(lr){3-5} \cmidrule(lr){6-8}
    Name &
    \multicolumn{1}{c}{State count} &
    \multicolumn{1}{c}{Total} & \multicolumn{1}{c}{Initial} & \multicolumn{1}{c}{Exploration} &
    \multicolumn{1}{c}{Total} & \multicolumn{1}{c}{Initial} & \multicolumn{1}{c}{Exploration}\\
    \midrule
$\memory_5$  & 88            & 301    & 301 & 0      & 448    & 448 & 0      \\
$\memory_6$  & 208           & 248    & 248 & 0      & 317    & 317 & 0      \\
$\memory_7$  & 480           & 248    & 248 & 0      & 319    & 319 & 0      \\
$\memory_8$  & 1,088         & 253    & 253 & 0      & 249    & 249 & 0      \\
$\memory_9$  & 2,432         & 250    & 246 & 3      & 321    & 317 & 4      \\
$\memory_{10}$ & 5,376         & 269    & 249 & 21     & 340    & 313 & 28     \\
$\memory_{11}$ & 11,776        & 286    & 257 & 29     & 360    & 314 & 45     \\
$\memory_{12}$ & 25,600        & 290    & 247 & 43     & 353    & 304 & 49     \\
$\memory_{13}$ & 55,296        & 297    & 247 & 51     & 362    & 294 & 68     \\
$\memory_{14}$ & 118,784       & 283    & 247 & 36     & 357    & 307 & 50     \\
$\memory_{15}$ & 253,952       & 282    & 244 & 38     & 354    & 300 & 55     \\
$\memory_{16}$ & 540,672       & 284    & 242 & 42     & 358    & 300 & 58     \\
$\memory_{17}$ & 1,146,880     & 289    & 243 & 46     & 372    & 304 & 68     \\
$\memory_{18}$ & 2,424,832     & 300    & 244 & 55     & 387    & 305 & 83     \\
$\memory_{19}$ & 5,111,808     & 351    & 254 & 97     & 471    & 339 & 132    \\
$\memory_{20}$ & 10,747,904    & 363    & 246 & 117    & 475    & 302 & 173    \\
$\memory_{21}$ & 22,544,384    & 423    & 250 & 173    & 560    & 302 & 257    \\
$\memory_{22}$ & 47,185,920    & 545    & 235 & 309    & 697    & 278 & 420    \\
$\memory_{23}$ & 98,566,144    & 869    & 239 & 630    & 1,061  & 302 & 759    \\
$\memory_{24}$ & 205,520,896   & 1,518  & 248 & 1,269  & 1,775  & 294 & 1,481  \\
$\memory_{25}$ & 427,819,008   & 2,771  & 249 & 2,522  & 3,389  & 299 & 3,090  \\
$\memory_{26}$ & 889,192,448   & 5,586  & 251 & 5,335  & 6,957  & 289 & 6,668  \\
$\memory_{27}$ & 1,845,493,760 & 14,791 & 251 & 14,539 & 16,892 & 307 & 16,584\\
    \bottomrule
\end{tabular}}
  \caption{Results of running \GPUexplore inclusion checking on $\memory_k$ instances.\label{table:inclusion-checking}}
  \end{table}


\section{Conclusions \& Future work}
\label{sec:conc}
We implemented and compared different parallel algorithms for DFA minimisation
on GPUs. We find that the $NC$ algorithm~$\texttt{trans}$ with parallel
logarithmic run-time does not scale well because of the large number of
resources needed. Instead, we find that the partition refinement algorithms
perform better. This might be seen as contradictory since these partition
refinement algorithms have inherently linear parallel run-times. 

When comparing the different partition refinement algorithms, the structure of the
input DFA is of high influence. The trade-off is that in $\texttt{sortPR}$ each
iteration takes more time than in $\texttt{naivePR}$, but in \texttt{sortPR},
an iteration has the potential to lead to a block
being split into more than two subblocks. When this happens sufficiently often,
fewer iterations are needed. This leads to
$\texttt{sortPR}$ being slower in cases where the number of iterations is high.
In other benchmarks, it leads to fewer iterations and thereby a significant
speed-up. 

Finally, we showed a way to incorporate a partial transitive closure in
partition refinement algorithms. We showed that for a specific class of DFAs
this approach leads to logarithmic run-times, where every partition refinement
algorithm is inherently linear.

Regarding equivalence and inclusion checking, we proposed a way to perform
a naive variant of the Hopcroft-Karp algorithm massively parallel on a GPU,
by encoding these problems in \SLCO, an input language of the
GPU-accelerated explicit-state model checking \GPUexplore. We presented
several benchmarks, and demonstrated that for multiple of these, an almost
linear increase of the overall execution time can be observed, meaning that
the lack of a union-find data structure, which is used in the original Hopcroft-Karp
algorithm, is largely mitigated by the use of parallelism.

As future work it would be interesting to further investigate sublinear time
parallel algorithms for DFA minimisation. Specifically, there are two key
questions that come to mind. The first question is: what is a reasonable number
of parallel processors necessary for a poly-logarithmic time parallel algorithm?
It seems feasible to use a similar argument as in~\cite{khuller1994parallel} to
get a superlinear lower bound. However, the gap between that and the $O(n^{2\omega})$
processors%
\footnote{If matrix multiplication can be computed in time $O(n^\omega)$,
currently best known bounds are $\omega \leq 2.372\dots$.} used in
Algorithm~\ref{alg:fulltrans} remains large. The second question is: can a method
such as the one presented in this article using the partial transitive closure be
implemented in such a way that the run
time will be sublinear with high probability, i.e. any parallel run-time
$O(n^{1{-}\epsilon})$ for some $\epsilon \geq 0$? A good starting point to answer this question
would be the recent work on parallel reachability
algorithms~\cite{ullman1990high,fimeman2018nearwork,jambulapati2019parallel}.

Interesting future work for equivalence and inclusion checking would be to consider
the union-find data structure used in the Hopcroft-Karp algorithm to achieve
a near-linear complexity. If a massively parallel version of such a data structure
can be designed, it can be expected that the execution times of our GPU
approach will improve. However, to what extent they will improve in practice will
require further experimentation.

\bibliographystyle{alphaurl}
\bibliography{main}
\end{document}